\documentclass[amsthm]{elsart}

\usepackage{ifpdf}
\usepackage{yjsco}
\usepackage{natbib}
\usepackage{bbm,mathrsfs}
\usepackage{amsmath}
\usepackage{alltt, amssymb}
\usepackage{graphicx}

\def\d {\ensuremath{\mathbf{d}}}
\def\e {\ensuremath{\mathbf{e}}}
\def\C {\ensuremath{\mathsf{C}}}
\def\L {\ensuremath{\mathsf{L}}}

\def\N {\ensuremath{\mathbb{N}}}

\def\S {\ensuremath{\mathbf{S}}}
\def\Z {\ensuremath{\mathbb{Z}}}
\def\F {\ensuremath{\mathbb{F}}}

\def\X {\ensuremath{\mathbf{X}}}
\def\T {\ensuremath{\mathbf{T}}}
\def\M {\ensuremath{\mathsf{M}}}

\def\U {\ensuremath{\mathbf{U}}}
\def\V {\ensuremath{\mathbf{V}}}

\def\rng {\ensuremath{\mathsf{R}}}

\def\Span {\ensuremath{\mathsf{Span}}}
\def\Eval {\ensuremath{\mathsf{Eval}}}
\def\Lift {\ensuremath{\mathsf{Lift}}}
\def\Interp {\ensuremath{\mathsf{Interp}}}

\theoremstyle{plain}

\newtheorem{Corollary}{Corollary}

\newtheorem{Proposition}{Proposition}
\newtheorem{Lemma}{Lemma}
\newtheorem{Remark}{Remark}
\newcommand{\ead}[1]{(\texttt{#1})}

\begin{document}
\begin{frontmatter}
\title{Homotopy techniques for multiplication modulo triangular sets}
\thanks{This research was supported by ANR Gecko, the joint
  Inria-Microsoft Research Lab, MITACS, NSERC and the Canada Research
  Chair program.}

\author{Alin Bostan} 
\ead{alin.bostan@inria.fr}
\address{Algorithms Project, INRIA Rocquencourt, 78153 Le Chesnay Cedex, France}
\author{Muhammad Chowdhury} 
\ead{mchowdh3@csd.uwo.ca}
\address{Computer Science Department, The University of Western Ontario, London, Ontario, Canada}
\author{Joris van der Hoeven} 
\ead{Joris.Vanderhoeven@math.u-psud.fr}
\address{CNRS, D\'epartement de Math\'ematiques, Universit\'e Paris-Sud, 91405 Orsay Cedex, France}
\author{{\'E}ric Schost} 
\ead{eschost@uwo.ca}
\address{Computer Science Department, The University of Western Ontario, London, Ontario, Canada}

\begin{abstract}
  We study the cost of multiplication modulo triangular families of
  polynomials. Following previous work by Li, Moreno Maza and Schost,
  we propose an algorithm that relies on homotopy and fast
  evaluation-interpolation techniques. We obtain a quasi-linear time
  complexity for substantial families of examples, for which no such
  result was known before. Applications are given to notably addition
  of algebraic numbers in small characteristic.
\end{abstract}

\maketitle
\end{frontmatter}

%%%%%%%%%%%%%%%%%%%%%%%%%%%%%%%%%%%%%%%%%%%%%%%%%%%%%%%%%%%%
%%%%%%%%%%%%%%%%%%%%%%%%%%%%%%%%%%%%%%%%%%%%%%%%%%%%%%%%%%%%
%%%%%%%%%%%%%%%%%%%%%%%%%%%%%%%%%%%%%%%%%%%%%%%%%%%%%%%%%%%%

\section{Introduction}\label{sec:intro}

Triangular families of polynomials are a versatile data structure,
well adapted to encode geometric problems with some form of
symmetry~\cite{AuVa00,KoMo02,FoMo02,GaSc04}. However, in spite of
this, many complexity questions are still not answered in a satisfying
manner.

A high-level question is to provide sharp estimates on the cost of
solving polynomial systems by means of triangular
representations. This problem has a geometric nature; it itself relies
on several difficult lower-level questions, such as the cost of basic
operations with triangular sets. In this paper, we address one such
question: the arithmetic cost of multiplication of polynomials modulo
a triangular set. This justifiably stands as a central question,
since many higher-level routines are built on top of it, such as
inversion~\cite{Langemyr91,HoMo02,LiMoSc07}, lifting
techniques~\cite{DaMaScWuXi05} for modular algorithms, solving systems
of equations~\cite{LiMoRaSc08b}, etc.

\paragraph*{Problem statement, overview of our results.}
We work in $\rng[X_1,\dots,X_n]$, where $\rng$ is a ring, and we are
given a set of relations of the form
$$\T~\left | \begin{array}{l}
T_n(X_1,\dots,X_n)\\
~~~\vdots\\
T_2(X_1,X_2)\\
T_1(X_1).
\end{array}\right .$$
The polynomials $\T$ form a {\em triangular set}: for all $i$, $T_i$
is in $\rng[X_1,\dots,X_i]$, is {\em monic} in $X_i$ and {\em reduced}
modulo $ \langle T_1,\dots,T_{i-1} \rangle$, in the sense that
$\deg(T_i,X_j) < \deg(T_j,X_j)$ for $j < i$. As an aside, note that
the case where $T_i$ is not monic but with a leading coefficient
invertible modulo $\langle T_1,\dots,T_{i-1} \rangle$ reduces in
principle to the monic case; however, inversion modulo $\langle
T_1,\dots,T_{i-1} \rangle$ remains a difficult
question~\cite{DaMoScXi06}, of complexity higher than that of
multiplication.

As input, we consider two polynomials $A,B$ reduced modulo $\langle
\T\rangle$. The direct approach to multiply them modulo $\langle
\T\rangle$ is to perform a polynomial multiplication, followed by the
reduction modulo $\langle \T\rangle$, by a generalization of Euclidean
division. As far as complexity is concerned, when the number of
variables grows, this kind of approach cannot give linear time
algorithms. Consider for instance the case where all $T_i$ have degree
2 in their main variables $X_i$. Then, $A$ and $B$ both have $2^n$
monomials, but their product before reduction has $3^n$ monomials;
after reduction, the number of monomials is $2^n$ again.  If we let
$\delta=2^n$ be a measure of the input and output size, the cost of
such an algorithm is at least $3^n=\delta^{\log_2 3}$.

In this paper, we show that a different approach can lead to a
quasi-linear time algorithm, in cases where the monomial support of
$\T$ is sparse, or when the polynomials in $\T$ have a low total
degree. 
This will for example be the case for systems of the form
\begin{equation}
  \label{eq:T1}
\left | \begin{array}{l}
X_n^2-2X_{n-1}\\
~~~\vdots\\
X_2^2-2X_1\\
X_1^2
\end{array}\right .
\quad\text{or}\quad
\left | \begin{array}{l}
X_n^2-X_{n-1}\\
~~~\vdots\\
X_2^2-X_1\\
X_1^2,
\end{array}\right .
\end{equation}
whose applications are described later on. Our result also applies to
the following construction: start from $F\in \rng[X]$, say
$F=X^3-X^2+X-3$, and define the so-called ``Cauchy
modules''~\cite{ReVa99}, which are in effective Galois
theory~\cite{ReVa99,AbOrReVa04,ReYo06}:
\begin{equation}
  \label{eq:T2}
\left | \begin{array}{lllll}
F_3(X_1,X_2,X_3)&=&\frac{F_2(X_1,X_2)-F(X_1,X_3)}{X_2-X_3}&=&X_3+X_2+X_1-1\\[1mm]
F_2(X_1,X_2)&=&\frac{F_1(X_1)-F_1(X_2)}{X_1-X_2} &=& X_2^2 + X_2X_1 - X_2 + X_1^2 - X_1 + 1\\[1mm]
F_1(X_1)&=&F(X_1)&=&X_1^3-X_1^2+X_1-3.
\end{array}\right .
\end{equation}
For examples~\eqref{eq:T1} and~\eqref{eq:T2}, our algorithms give the
following results:

\smallskip

\begin{itemize}
\item for $\T$ as in~\eqref{eq:T1}, multiplication modulo $\langle \T
  \rangle$ can be performed in quasi-linear time $O\tilde{~}(\delta)$,
  where $\delta=2^n$ is the input and output size,
  and where $O\tilde{~}(\delta)$ stands for $O(\delta (\log \delta)^{O(1)})$.

\smallskip

\item for $\T$ as in~\eqref{eq:T2}, with $n=\deg(F)$, multiplication
  modulo $\langle \T \rangle$ can be performed in quasi-linear time
  $O\tilde{~}(\delta)$, where $\delta=n!$ is the input and output
  size.
\end{itemize}

\smallskip\noindent
No previous algorithm was known featuring such complexity estimates.

\paragraph*{Our approach.}
To obtain this quasi-linear cost, we have to avoid multiplying $A$ and
$B$ as polynomials. Our solution is to use evaluation and
interpolation techniques, just as FFT multiplication of univariate
polynomials is multiplication modulo $X^n-1$.

Fast evaluation and interpolation may not be possible directly, if
$\T$ does not have roots in $\rng$ (as in the previous
examples). However, they become possible using deformation techniques:
we construct a new triangular set $\U$ with all roots in $\rng$, and
multiply $A$ and $B$ modulo $\S=\eta \T + (1-\eta) \U$, where $\eta$
is a new variable. The triangular set $\S$ has roots in
$\rng[[\eta]]$, by Hensel's lemma, so one can use evaluation-interpolation
techniques over $\rng[[\eta]]$.

This idea was introduced in~\cite{LiMoSc07}, but was limited to the
case where all polynomials in $\T$ are univariate: $T_i$ was
restricted to depend on $X_i$ only, so this did not apply to the
examples above. Here, we extend this idea to cover such examples; our
main technical contribution is a study of precision-related issues
involved in the power series computations, and how they relate to the
monomial support of $\T$.

\paragraph*{Previous work.} It is only recently that fast algorithms
for triangular representations have been throughly investigated; thus,
previous results on efficient multiplication algorithms are scarce.
All natural approaches introduce in their cost estimate an overhead of
the form $k^n$, for some constant $k$. 

The main challenge (still open) is to get rid of this exponential
factor unconditionnally: we want algorithms of cost
$O\tilde{~}(\delta)$, where $\delta=d_1\cdots d_n$ is the number of
monomials in $A$, $B$ and their product modulo $\langle \T \rangle$.
For instance, with $T_i=X_i^{d_i}$, the first complexity result of the form
$O(\delta^{1+\varepsilon})$, for any $\varepsilon > 0$, was in~\cite{Schost05}.

The previous work~\cite{LiMoSc07} gives a general algorithm of cost
$O\tilde{~}(4^n\delta)$. That algorithm uses fast Euclidean division;
for polynomials of low degree (e.g., for $d_i=2$), the naive approach
for Euclidean division can actually give better results.  Previous
mentions of such complexity estimates (with a constant higher than
$4$) are in~\cite{Langemyr91}.

As said above, in~\cite{LiMoSc07}, one also finds the precursor of the
algorithm presented here; the algorithm of~\cite{LiMoSc07} applies to
families of polynomials having $T_i \in \rng[X_i]$, and achieves the
cost $O(\delta^{1+\varepsilon})$ for any $\varepsilon > 0$. In that
case, the analysis of the precision in power series computation was
immediate. Our main contribution here is to perform this study in the
general case, and to show that we can still achieve similar costs for
much larger families of examples.

\paragraph*{Basic notation.}
Let $\d=(d_1,\dots,d_n)$ be a vector of positive integers. In what
follows, these will represent the main degrees of the polynomials in
our triangular sets; without loss of generality, we will thus always
suppose $d_i \ge 2$ for all $i$. 

Recall that $\rng$ is our base ring and that $X_1,\dots,X_n$ are
interminates over $\rng$. We let $M_\d$ be the set of monomials
$$M_\d= \big \{ X_1^{e_1}\cdots X_n^{e_n} \ | \ 0 \leq e_i < d_i
\text{~for~all~} i\ \big \}.$$ We
denote by $\Span(M_\d)$ the free $\rng$-submodule generated by $M_\d$
in $\rng[X_1,\dots,X_n]$:
$$\Span(M_\d) = \big \{\ A=\sum_{m \in M_\d} a_m m \ | \ a_m \in \rng \text{~for all $m \in M_\d$}\ \big
 \}.$$
This is thus the set of polynomials $A$ in $\rng[X_1,\dots,X_n]$, 
such that $\deg(A_i,X_i) < d_i$ holds for all $i$.
Finally, we let $\delta_\d$ be the product $\delta_\d=d_1\cdots d_n$;
this is the cardinality of $M_\d$. Remark that since all $d_i$ are at
least 2, we have the bounds
$$2^n \le \delta_\d \quad\text{and}\quad\sum_{i \le n} d_1\cdots d_i \le 2
\delta_\d.$$ The former plainly follows from the inequality $2 \le
d_i$; the latter comes from observing that $d_1\cdots d_i 2^{n-i} \le
d_1\cdots d_n = \delta_\d$; this yields $d_1\cdots d_i \le \delta_\d /
2^{n-i}$, from which the claim follows by summation.

The {\em multi-degree} of a triangular set $\T=(T_1,\dots,T_n)$ is the
$n$-uple $\d=(d_1,\dots,d_n)$, with $d_i=\deg(T_i,X_i)_{1 \leq i\leq
  n}$. In this case, $\rng[X_1,\dots,X_n]/\langle \T \rangle$ is a
free $\rng$-module isomorphic to $\Span(M_\d)$.  We say that a
polynomial $A \in \rng[X_1,\dots,X_n]$ is {\em reduced} with respect
to $\T$ if $\deg(A,X_i) < d_i$ holds for all $i$. For any $A \in
\rng[X_1,\dots,X_n]$, there exists a unique $A' \in
\rng[X_1,\dots,X_n]$, reduced with respect to $\T$, and such that
$A-A'$ in is the ideal $\langle \T\rangle$. We call it the {\em normal
  form} of $A$ and write $A' = A \bmod \langle \T \rangle$.

\paragraph*{Outlook of the paper.} In Section~\ref{sec:prel}, we
introduce some basic complexity notation. The next section presents
basic evaluation-interpolation algorithms for so-called {\em
  equiprojectable} sets, which are extensions of algorithms known for
univariate polynomials. We deduce our multiplication algorithm in
Section~\ref{sec:mul}; examples, applications and experimental results
are in Sections~\ref{sec:appli} and~\ref{sec:algnum}.

%%%%%%%%%%%%%%%%%%%%%%%%%%%%%%%%%%%%%%%%%%%%%%%%%%%%%%%%%%%%
%%%%%%%%%%%%%%%%%%%%%%%%%%%%%%%%%%%%%%%%%%%%%%%%%%%%%%%%%%%%
%%%%%%%%%%%%%%%%%%%%%%%%%%%%%%%%%%%%%%%%%%%%%%%%%%%%%%%%%%%%

\section{Preliminaries}\label{sec:prel}

Big-O notation is delicate to use in our situation, since our
estimates may depend on several (possibly an unbounded number of)
parameters (typically, the multi-degree of our triangular
sets). Hence, whenever we use a big-O inequality such as $f \in O(g)$,
it is implied that there exists a universal constant $\lambda$ such
that $f(v_1,\dots,v_s) \le \lambda g(v_1,\dots,v_s)$ holds for {\em
  all} possible values of the arguments. When needed, we use explicit
inequalities. Finally, the notation $f \in O\tilde{~}(g)$ means that
there exists a constant $\alpha$ such that $f \in O(g
\log(g)^\alpha)$, where the big-O is to be understood as above.

Our complexity estimates count additions, multiplications, and
inversions, when they are possible. We denote by $\M:\N \to\N$ a
function such that over any ring, polynomials of degree less than $d$
can be multiplied in $\M(d)$ operations, and which
satisfies the super-linearity conditions of~\cite[Chapter
8]{GaGe99}. Using the algorithm of Cantor-Kaltofen~\cite{CaKa91}, one
can take $\M(d) \in O(d \lg(d)\lg\lg(d))$, with $\lg(d)=\log_2
\max(d,2)$.

We next let $\C_0:\N \to \N$ be a function such that $\C_0(d) \ge d$
holds for all $d$ and such that we have, over any ring $\rng$:
\begin{enumerate}
\item for any $x_1,\dots,x_d$ in $\rng$ and any polynomial $A \in
  \rng[X]$ of degree less than $d$, one can compute all values
  $A(x_i)$ in $\C_0(d)$ additions and multiplications in $\rng$;

\smallskip

\item for any $x_1,\dots,x_d$ in $\rng$, one can compute the
  coefficients of the polynomial $(X-x_1)\cdots (X-x_d)$ in $\C_0(d)$
  additions and multiplications in $\rng$;

\smallskip

\item for any $x_1,\dots,x_d$ in $\rng$, with $x_i - x_j$ a unit for
  $i\ne j$, and any values $v_1,\dots,v_d$ in $\rng$, one can compute
  the unique polynomial $A \in \rng[X]$ of degree less than $d$ such
  that $A(x_i)=v_i$ holds for all $i$ in $\C_0(d)$ operations in
  $\rng$.
\end{enumerate}

\smallskip\noindent
By the results of~\cite[Chapter 10]{GaGe99}, one can take $\C_0(d) \in
O(\M(d)\log(d))$. We continue with the well-known fact that the function
$\C_0$ also enables us to estimate the cost of lifting power series
roots of a bivariate polynomial by Newton iteration. In the following
lemma, $\eta$ is a new variable over $\rng$.

\begin{Lemma}\label{lemma:lift}
  For any polynomial $T$ in $\rng[\eta,X]$, monic in $X$ and with
  $\deg(T,X)=d$, if the roots of $T(0,X)$ are known and have
  multiplicity 1, one can compute the roots of $T(\eta,X)$ in
  $\rng[\eta]/\langle \eta^\ell\rangle$ in $O(\C_0(d)\M(\ell))$
  operations in $\rng$.
\end{Lemma}

\begin{pf}
  The algorithm consists in lifting all roots of $T$ in parallel using
  Newton iteration, using fast evaluation to compute the needed values
  of $T$ and $\partial T/ \partial X$; it is given in
  Figure~\ref{Fig:1}, where we use a subroutine called ${\sf
    EvalUnivariate}$ to do the evaluation.  Each pass through the loop
  at line 2 takes two evaluations in degree $d$ and $d$ inversions,
  with coefficients that are power series of precision
  $\ell'$. Using~\cite[Chapter~9]{GaGe99}, the cost is thus $2 \C_0(d)
  \M(\ell') + \lambda d \M(\ell')$, for some constant $\lambda$. Using
  the super-linearity of the function $\M$, the conclusion follows.
\end{pf}

\begin{Remark}\textup{
  When performing all multiplications in
  $\rng[\eta,X]/\langle \eta^\ell\rangle$ using Kronecker's method,
  a more precise cost analysis yields the bound $O(\M(d \ell) \log(d))$
  instead of
  $$O(\C_0(d) \M(\ell)) = O(\M(d) \M(\ell) \log d).$$
  If $\ell=O(d)$, then we usually have $\M(d \ell) = O(\M(d) \ell)$,
  which makes the new bound slightly better.
  However, this improvement only has a minor impact on what follows,
  so it will be more convenient to use the technically simpler bound
  from the lemma.}
\end{Remark}

\begin{figure}[!!!h]
\begin{center}
\fbox{
\begin{minipage}{5 cm}
\begin{tabbing}
\qquad \= \qquad \= \quad \= \quad \kill
$\underline{{\sf LiftRoots}(T,a_1,\dots,a_d,\ell)}$\\[1mm]
1 \> $\ell' \leftarrow 2$\\
2 \> while $\ell' < \ell$ do\\
2.1 \> \> $v_1,\dots,v_d \leftarrow {\sf EvalUnivariate}(T, a_1,\dots,a_d) \bmod \eta^{\ell'}$\\
2.2 \> \> $w_1,\dots,w_d \leftarrow {\sf EvalUnivariate}(\partial T/\partial X, a_1,\dots,a_d) \bmod \eta^{\ell'}$\\
2.3 \> \> for $i=1,\dots,d$ do \\
2.3.1 \> \> \> $a_i \leftarrow a_i - v_i/w_i \bmod \eta^{\ell'}$\\
2.4 \> \> $\ell' \leftarrow 2 \ell'$\\
3 \> return $[a_i \bmod X^{\ell} \ | \ 1 \le i \le d]$
\end{tabbing}
\end{minipage}
}
\caption{Lifting all roots of a univariate polynomial.}
\label{Fig:1}
\end{center}
\end{figure}

To obtain simpler estimates, we let $\C(d)= \Lambda \C_0(d)$, where
$\Lambda\ge 1$ is the constant implied in the big-O estimate in the
former lemma. Hence, problems (1), (2) and (3) above can be dealt with
in $\C(d)$ operations, and the lifting problem of the previous lemma
can be solved in $\C(d)\M(\ell)$ operations.  Finally, we introduce
another short-hand notation: for a multi-degree $\d=(d_1,\dots,d_n)$,
we write
\begin{equation}
  \label{eq:L}
\L(\d) =\sum_{i \le n} \frac {\C(d_i)}{d_i} \le n  \frac{\C(d)}{d},  
\end{equation}
with $d=\max_{i \le n} d_i$. In view of the estimates on $\C$, we
also have the upper bound $\L(\d) \in O(\lg(\delta_\d)^3)$, which shows
that $\L(\d)$ is of polylogarithmic growth in $\delta_\d$.

%%%%%%%%%%%%%%%%%%%%%%%%%%%%%%%%%%%%%%%%%%%%%%%%%%%%%%%%%%%%
%%%%%%%%%%%%%%%%%%%%%%%%%%%%%%%%%%%%%%%%%%%%%%%%%%%%%%%%%%%%
%%%%%%%%%%%%%%%%%%%%%%%%%%%%%%%%%%%%%%%%%%%%%%%%%%%%%%%%%%%%

\section{Evaluation and interpolation at equiprojectable sets}\label{sec:equi}

In this section, we recall from~\cite{AuVa00} the definition of {\em
  equiprojectable sets}.  We prove that one can perform evaluation and
interpolation at, and construct the vanishing ideal of,
equiprojectable sets in linear time, up to logarithmic factors. We
deduce an algorithm for multiplication modulo the vanishing ideal of
such sets with a similar complexity. These results extend those given
in~\cite{Pan94,LiMoSc07}, which dealt with the case of points on a
regular grid. The extension to our more general context is rather
straightforward, but to our knowledge, it has not appeared in print
before.

In all this section, $\rng$ is a ring; we study subsets of $\rng^n$
and their successive projections on the subspaces $\rng^i$, for $i\le
n$. For definiteness, we let $\rng^0$ be a one-point set.  Then, for $1
\le j \le i \le n$, we let $\pi_{i,j}$ be the projection
$$\begin{array}{cccc}
\pi_{i,j}:& \rng^i & \to & \rng^{j} \\
& (x_1,\dots,x_i) & \mapsto & (x_1,\dots,x_j);
\end{array}$$
if $j=0$, we adapt this definition by letting $\pi_{i,0}$ be the
constant map $\rng^i \to \rng^0$. Finally, since this is the
projection we use most, we simply write $\pi=\pi_{n,n-1}$ for the
projection $\rng^n \to \rng^{n-1}$.

If $V$ is a subset of $\rng^n$, for $\beta$ in $\pi(V)$, we let
$V_\beta$ be the fiber $V \cap \pi^{-1}(\beta)$. Hence, if $\beta$ has
coordinates $(\beta_1,\dots,\beta_{n-1})$, the points in $V_\beta$
have the form $(\beta_1,\dots,\beta_{n-1},a)$, for some values $a$ in
$\rng$.  In all that follows, a finite set is by convention non-empty.

%%%%%%%%%%%%%%%%%%%%%%%%%%%%%%%%%%%%%%%%%%%%%%%%%%%%%%%%%%%%

\subsection{Equiprojectable sets}

Let $V$ be a finite set in $\rng^n$. {\em Equiprojectability} is a
property of $V$ that describes a combinatorial regularity in the
successive projections of $V$. For $n=0$, we say that the unique
non-empty subset of $\rng^0$ is equiprojectable. Then, for $n>0$,
$V\subset\rng^n$ is equiprojectable if the following holds:

\smallskip

\begin{itemize}
\item the projection $\pi(V)$ is equiprojectable in $\rng^{n-1}$, and

\smallskip

\item there exists an integer $d_n$ such that for all $\beta$ in
  $\pi(V)$, the fiber $V_\beta$ has cardinality $d_n$.
\end{itemize}

\smallskip\noindent
The vector $\d=(d_1,\dots,d_n)$ is called the {\em multi-degree} of
$V$. Remark that for $n=1$, any finite $V \subset \rng$ is equiprojectable.
One easily sees that if $V$ is equiprojectable, its cardinality equals
$\delta_\d=d_1\cdots d_n$; more generally, $\pi_{n,i}(V) \subset \rng^i$ is
equiprojectable of cardinality $d_1 \cdots d_i$. When $\rng$ is a
perfect field, it is proved in~\cite{AuVa00} that equiprojectable sets
are exactly the zero-sets of triangular sets that generate radical
ideals in $\rng[X_1,\dots,X_n]$; we will discuss this in more detail
in Subsection~\ref{sec:ideal}.

We give first a slightly more precise notation for the fibers $V_\beta$:
if $V$ is equiprojectable, then for all
$\beta=(\beta_1,\dots,\beta_{n-1}) \in \pi(V)$,
there exist exactly $d_n$ pairwise distinct values
$v_{\beta}=[a_{\beta,1},\dots,a_{\beta,d_n}]$ in $\rng$ such that
$$V_\beta = [ (\beta_1,\dots,\beta_{n-1},a_{\beta,i}) \ | \ a_{\beta,i} \in v_\beta ]$$
and thus
$$V\ =\ 
[ (\beta_1,\dots,\beta_{n-1},a_{\beta,i}) \ | \
\beta=(\beta_1,\dots,\beta_{n-1}) \in \pi(V),\ a_{\beta,i} \in
v_\beta,\ 1 \le i \le d_n ].$$ For instance, $n$-dimensional grids are
special cases of equiprojectable sets, where $v_{\beta}$ is
independent of $\beta$. Remark also that for some special choices of
$v_\beta$, improvements in the algorithms below are possible (e.g.,
for $v_\beta$ translates of points in geometric progression, using the
algorithm of~\cite{AhStUl75}).

%%%%%%%%%%%%%%%%%%%%%%%%%%%%%%%%%%%%%%%%%%%%%%%%%%%%%%%%%%%%

\subsection{Evaluation}

Let $V$ be an equiprojectable set of multi-degree $\d=(d_1,\dots,d_n)$,
and let $M_\d$, $\delta_\d$ be as in Section~\ref{sec:intro}. We
denote by $\Eval_V$ the evaluation map
$$\begin{array}{cccc}
\Eval_V:&\Span(M_\d) & \to & \rng^{\delta_\d} \\
& F & \mapsto &[F(\alpha) \ | \ \alpha \in V].
\end{array}$$
We let $\C_\Eval$ be a function such that for any $V$ equiprojectable
of multidegree $\d$, the map $\Eval_V$ can be evaluated in
$\C_\Eval(\d)$  operations. In one variable, with
$n=1$ and $\d=(d_1)$, $\C_\Eval(\d)$ simply describes the cost of evaluating a
polynomial of degree less than $d_1$ at $d_1$ points of $\rng$, so
we can take $\C_\Eval(\d) = \C(d_1)$. More generally, we have the following
quasi-linear time estimate.

\begin{Proposition}\label{prop:eval}
One can take $\C_\Eval(\d) \le \delta_\d \L(\d).$
\end{Proposition}

\begin{pf}
We will use a straightforward recursion over the variables $X_n,\ldots,X_1$.
Let $W =\pi(V)$, let $\e=(d_1,\dots,d_{n-1})$ be the
multi-degree of $W$ and let $A(X_1,\dots,X_n)\in\Span(M_\d)$ be
the polynomial to evaluate. We write $$A = \sum_{i < d_n}
A_i(X_1,\dots,X_{n-1}) X_n^i,$$ with $A_i$ in $\Span(M_{\e})$, and,
for $\beta$ in $\rng^{n-1}$, we define $$A_\beta = \sum_{i < d_n}
A_i(\beta) X_n^i\ \in\ \rng[X_n].$$ Hence, for
$\beta=(\beta_1,\dots,\beta_{n-1})$ in $\rng^{n-1}$ and $x$ in $\rng$,
$A(\beta_1,\dots,\beta_{n-1},x) = A_\beta(x)$. As a consequence, to
evaluate $A$ at $V$, we start by evaluating all $A_i$ at all points
$\beta\in W$. This gives all polynomials $A_\beta$, which we evaluate
at the fibers $v_\beta$.
\begin{figure}[!!!h]
\begin{center}
\fbox{
\begin{minipage}{5 cm}
\begin{tabbing}
\qquad \= \qquad \= \quad \= \quad \kill
$\underline{\Eval(A,V)}$\\[1mm]
1 \> if $n = 0$ return $[A]$\\
2 \> $W \leftarrow \pi(V)$\\
3 \> for $i =0,\dots,d_n-1$ do\\
3.1 \> \> $A_i \leftarrow {\sf coeff}(A, X_n, i)$ \\
3.2 \> \> ${\rm val}[i] \leftarrow \Eval(A_i, W)$ \\
\> \> $\slash\hspace{-1mm}*\ {\rm val}[i]$ has the form $[A_i(\beta) \ | \ \beta \in W]\ *\hspace{-1mm}\slash$\\
4 \> for $\beta$ in $W$ do \\
4.1 \> \> $A_\beta \leftarrow \sum_{i < d_n} A_i(\beta) X_n^i$\\
5 \> return $[{\sf EvalUnivariate}(A_\beta, v_\beta)\ |\ \beta \in W]$
\end{tabbing}
\end{minipage}
}
\caption{Evaluation algorithm.}
\label{Fig:2_1}
\end{center}
\end{figure}
The algorithm is given in Figure~\ref{Fig:2_1}. From this, we deduce
that we can take $\C_\Eval$ satisfying the recurrence
\begin{equation*}\C_\Eval(d_1,\dots,d_n) \leq 
     \C_\Eval(d_1,\dots,d_{n-1})\, d_n+ 
       \, d_1\cdots d_{n-1} \C(d_n).
\end{equation*}
This implies 
$$\C_\Eval(d_1,\dots,d_n)\ \le\ \sum_{i \le n} \delta_\d \frac{\C(d_i)}{d_i},$$
which proves the proposition. 
\end{pf}

%%%%%%%%%%%%%%%%%%%%%%%%%%%%%%%%%%%%%%%%%%%%%%%%%%%%%%%%%%%% 

\subsection{Interpolation}

Using the same notation as above, the inverse of the evaluation map is
interpolation at $V$:
$$\begin{array}{cccc}
\Interp_V:&\rng^{\delta_\d} & \to & \Span(M_\d) \\
& [F(\alpha) \ | \ \alpha \in V] & \mapsto & F.
\end{array}$$
For this map to be well-defined, we impose a natural condition on the
points of $V$. Let $W=\pi(V) \in \rng^{n-1}$. We say that $V$ {\em supports interpolation} if

\smallskip

\begin{itemize}
\item if $n > 1$, $W$ supports interpolation, and

\smallskip

\item for all $\beta$ in $W$ and all $x,x'$ in $v_\beta$, 
  $x-x'$ is a unit;
\end{itemize}

\smallskip
\noindent
if the base ring is a field, this condition is vacuous.  We will see
in the following proposition that if $V$ supports interpolation,
then the map $\Interp_V$ is well-defined. Moreover, we let $\C_\Interp$
be such that, for $V$ equiprojectable of multi-degree $\d$,
if $V$ supports interpolation, then the map $\Interp_V$
can be evaluated in $\C_\Interp(\d)$ operations (including inversions).
\begin{Proposition}\label{prop:interp}
  If $V$ supports interpolation, the map $\Interp_V$ is well-defined.
  Besides, one can take $\C_\Interp(\d) \le \delta_\d \L(\d).$
\end{Proposition}
\begin{pf}
  If $n=0$, we do nothing; otherwise, we let $W=\pi(V)$. The set of
  values to interpolate at $V$ has the shape $[f_\alpha \ | \ \alpha
  \in V] \in \rng^{\delta_\d}$; we can thus rewrite it as $[f_\beta \
  |\ \beta \in W]$, where each $f_\beta$ is in~$\rng^{d_n}$.

Since $V$ supports interpolation, for $\beta$ in $W$, there exists a
unique polynomial $A_\beta\in \rng[X_n]$ of degree less than $d_n$,
such that $\Eval(A_\beta, v_\beta)=f_\beta$. Applying the
algorithm recursively on the coefficients of the polynomials $A_\beta$,
we can find a polynomial $A$ such that $A(\beta,X_n)=A_\beta(X_n)$
holds for all $\beta \in W$.  Then, the polynomial $A$ satisfies our
constraints. This provides a right-inverse, and thus a two-sided
inverse for the map $\Eval$.
\begin{figure}[!!!h]
\begin{center}
\fbox{
\begin{minipage}{5 cm}
\begin{tabbing}
\qquad \= \qquad \= \quad \= \quad \kill
$\underline{\Interp(f,V)}$\\[1mm]
1 \> if $n = 0$ return $[f]$\\
2 \> $W \leftarrow \pi(V)$\\
3 \> for $\beta$ in $W$ do \\
3.1 \> \> $A_\beta \leftarrow {\sf InterpUnivariate}(f_\beta,v_\beta)$\\
4 \> for $i =0,\dots,d_n-1$ do\\
4.1 \> \> $c_i \leftarrow [{\sf coeff}(A_\beta, X_n, i) \ | \ \beta \in W]$ \\
4.2 \> \> $A_i \leftarrow \Interp(c_i, W)$\\
5 \> return $\sum_{i < d_n} A_i X_n^i$
\end{tabbing}
\end{minipage}
}
\caption{Interpolation algorithm.}
\label{Fig:2_2}
\end{center}
\end{figure}
The algorithm is given in Figure~\ref{Fig:2_2}; we use a
subroutine called {\sf InterpUnivariate} for univariate
interpolation. As for evaluation, we deduce that we can take
$\C_\Interp$ satisfying
\begin{equation*}\
\C_\Interp(d_1,\dots,d_n) \leq 
     \C_\Interp(d_1,\dots,d_{n-1})\, d_n+ 
       \, d_1\cdots d_{n-1} \C(d_n),
\end{equation*}
which gives our claim, as in the case of evaluation.\end{pf}

%%%%%%%%%%%%%%%%%%%%%%%%%%%%%%%%%%%%%%%%%%%%%%%%%%%%%%%%%%%%

\subsection{Associated triangular set}\label{sec:ideal}

Next, we associate to an equiprojectable set $V \subset \rng^n$ of
multi-degree $\d=(d_1,\dots,d_n)$ a triangular set
$\T=(T_1,\dots,T_n)$ of the same multi-degree, which vanishes on
$V$. As soon as $V$ supports interpolation, the existence of $\T$ is
guaranteed (and is established in the proof of the next
proposition). Uniqueness holds as well: if $(T_1,\dots,T_n)$ and
$(T'_1,\dots,T'_n)$ both vanish on $V$ and have multi-degree $\d$,
then for all $i$, $T_i-T'_i$ vanishes at $V$ as well and is in
$\Span(M_\d)$; hence, it is zero.  We call $\T$ the {\em associated
  triangular set}; if $\rng$ is a field, $\T$ is a lexicographic
Gr{\"o}bner basis of the vanishing ideal of $V$.

\begin{Proposition}\label{prop:van}
  Given an equiprojectable set $V$ of multi-degree $\d$ that supports
  interpolation, one can construct the associated triangular set $\T$
  in time $O(\delta_\d \L(\d))$.
\end{Proposition}
\begin{pf} We proceed inductively, and suppose that we already have
  computed $T_1,\dots,T_{n-1}$ as the associated triangular set of
  $W=\pi(V)$. We will write $\d=(d_1,\dots,d_n)$ and
  $\e=(d_1,\dots,d_{n-1})$.

  For $\beta$ in $W$, let $T_{\beta}$ be the polynomial $\prod_{a \in
    v_\beta} (X_n-a) \in \rng[X_n]$. For $j < d_n$, let further
  $T_{j,n}$ be the polynomial in $\Span(M_\e)$ that interpolates the
  $j$th coefficient of the polynomials $T_{\beta}$ at $W$; for
  $j=d_n$, we take $T_{d_n,n}=1$. We then write $T_n=\sum_{j \le d_n}
  T_{j,n} X_n^j$: this polynomial is in $\rng[X_1,\dots,X_n]$, monic
  of degree $d_n$ in $X_n$, has degree less than $d_i$ in $X_i$, for
  $i<n$, and vanishes on $V$. Thus, the polynomials
  $\T=(T_1,\dots,T_n)$ form the triangular set we are looking for. The
  algorithm is in Figure~\ref{Fig:VI}; we use a function {\sf
    PolyFromRoots} to compute the polynomials $T_\beta$.

\begin{figure}[!!!h]
\begin{center}
\fbox{
\begin{minipage}{5 cm}
\begin{tabbing}
\qquad \= \qquad \= \quad \= \quad \kill
$\underline{{\sf AssociatedTriangularSet}(V,n)}$\\[1mm]
1 \> if $n = 0$ return $[]$\\
2 \> $W \leftarrow \pi(V)$\\
3 \> $(T_1,\dots,T_{n-1})\leftarrow {\sf AssociatedTriangularSet}(W,n)$ \\  
4 \> for $\beta$ in $W$ do\\
4.1 \> \> $T_\beta \leftarrow {\sf PolyFromRoots}(v_\beta)$\\
5 \> for $j=0,\dots,d_n-1$ do \\
5.1 \> \> $T_{j,n} \leftarrow \Interp([{\rm coeff}(T_\beta,X_n,j) \ | \ \beta \in W], W)$\\
6 \> return $\sum_{j < d_n} T_{j,n}X_n^j+X_n^{d_n}$
\end{tabbing}
\end{minipage}
}
\caption{Associated triangular set of $V$.}
\label{Fig:VI}
\end{center}
\end{figure}

For a given $\beta$ in $W$, the function {\sf PolyFromRoots} computes
$T_{\beta}$ in $\C(d_n)$ base ring operations; this implies that given
$T_1,\dots,T_{n-1}$, one can construct $T_n$ using $d_1\cdots d_{n-1}
\C(d_n) + \C_\Interp(d_1,\cdots,d_{n-1}) d_n$ operations. The total
cost for constructing all $T_i$ is thus at most
$$\sum_{i \le n} d_1\cdots d_{i-1} \C(d_i) + \sum_{i \le
n}\C_\Interp(d_1,\cdots,d_{i-1}) d_i.$$ Using the trivial bound $d_1
\cdots d_i \le \delta_\d$ for the left-hand term, and the bound given
in Proposition~\ref{prop:interp} for the right-hand one, we get the
upper bounds
$$\delta_\d\sum_{i \le n}  \frac{\C(d_i)}{d_i} + \sum_{i \le
n} d_1 \cdots d_i \sum_{j \le i-1} \frac {\C(d_j)}{d_j}\ \le \
\delta_\d\sum_{i \le n}  \frac{\C(d_i)}{d_i} + \sum_{i \le
n} d_1 \cdots d_i  \sum_{j \le n} \frac {\C(d_j)}{d_j}.
$$
Using the upper bound $\sum_{i \le n} d_1\cdots d_i \le 2 \delta_\d$,
we finally obtain the estimate $3 \delta_\d \L(\d)$.
\end{pf}

%%%%%%%%%%%%%%%%%%%%%%%%%%%%%%%%%%%%%%%%%%%%%%%%%%%%%%%%%%%%

\subsection{Multiplication}\label{ssec:mult}

Using our evaluation and interpolation algorithms, it becomes
immediate to perform multiplication modulo a triangular set $\T$
associated to an equiprojectable set. 
\begin{Proposition}\label{prop:mul}
  Let $V \subset \rng^n$ be an equiprojectable set of multi-degree
  $\d=(d_1,\dots,d_n)$ that supports interpolation, and let $\T$ be
  the associated triangular set. Then one can perform multiplication
  modulo $\langle \T\rangle$ in time $O(\delta_\d \L(\d))$.
\end{Proposition}
\begin{pf}
  The algorithm is the same as in~\cite[Section~2.2]{LiMoSc07}, except
  that we now use the more general evaluation and interpolation
  algorithms presented here. Let $A$ and $B$ be reduced modulo
  $\langle \T \rangle$, and let $C=AB \bmod \langle \T \rangle$. Then
  for all $\alpha$ in $V$, $C(\alpha)=A(\alpha)B(\alpha)$. Since $C$
  is reduced modulo $\langle \T \rangle$, it suffices to interpolate
  the values $A(\alpha)B(\alpha)$ to obtain $C$.
The cost is thus that of two evaluations, one interpolation, and of
all pairwise pairwise products; the bounds of Propositions~\ref{prop:eval}
and ~\ref{prop:interp} conclude the proof.
\end{pf}

\begin{figure}[!!!h]
\begin{center}
\fbox{
\begin{minipage}{5 cm}
\begin{tabbing}
\qquad \= \qquad \= \quad \= \quad \kill
$\underline{{\sf Mul}(A, B, V)}$\\[1mm]
1   \> ${\sf Val}_A \leftarrow \Eval(A,V)$\\
2   \> ${\sf Val}_B \leftarrow \Eval(B,V)$\\
3   \> ${\sf Val}_C \leftarrow [\ {\sf Val}_A(\alpha) {\sf Val}_B(\alpha) \ | \ \alpha \in V\ ]$\\
4   \> {\sf return} $\Interp({\sf Val}_C,V)$
\end{tabbing}
\end{minipage}
}
\caption{Multiplication algorithm.}
\label{Fig:2_3}
\end{center}
\end{figure}

%%%%%%%%%%%%%%%%%%%%%%%%%%%%%%%%%%%%%%%%%%%%%%%%%%%%%%%%%%%%
%%%%%%%%%%%%%%%%%%%%%%%%%%%%%%%%%%%%%%%%%%%%%%%%%%%%%%%%%%%%
%%%%%%%%%%%%%%%%%%%%%%%%%%%%%%%%%%%%%%%%%%%%%%%%%%%%%%%%%%%%

\section{Homotopy techniques for multiplication}\label{sec:mul}

Let $\T$ be a triangular set in $\rng[X_1,\dots,X_n]$. We saw in the
previous section that if $\T$ has all its roots in $\rng$, and if
$V(\T)$ supports interpolation, then multiplication modulo $\langle \T
\rangle$ can be done in quasi-linear time. In this section, we extend
this approach to an arbitrary $\T$ by setting up an homotopy between
$\T$ and a new, more convenient, triangular set $\U$. This extends the
approach of~\cite[Section~2.2]{LiMoSc07}, which dealt with the case
where $T_i$ is in $\rng[X_i]$ for all $i$.

Let $\d$ be the multi-degree of $\T$ and assume that there exists an
equiprojectable set $V$ in $\rng^n$ which supports interpolation and
has multi-degree $\d$. Let $\U$ be the triangular set associated
to $V$ and let $\eta$ be a new variable. We then define the set $\S$
in $\rng[[\eta]][X_1,\dots,X_n]$ by
$$S_i = \eta T_i + (1-\eta) U_i, \quad 1 \le i \le n.$$
Since $\U$ and $\T$ have the same multi-degree $\d$,
this set $\S$ is triangular, with multi-degree~$\d$. 

In Subsection~\ref{ssec:roots}, we prove that $\S$ has all its roots
in $\rng[[\eta]]$. Thus, we can use evaluation-interpolation techniques
to do multiplication modulo $\langle \S \rangle$; this will in turn be
used to perform multiplication modulo $\langle \T \rangle$. 

The algorithm involves computing with power series; the quantity that
will determine the cost of the algorithm will be the required
precision in $\eta$. For $\e=(e_1,\dots,e_n)$ in $\N^n$, we define
$$H_0(e_1,\dots,e_n) = \deg(X_1^{e_1}\cdots X_n^{e_n} \bmod \langle \S \rangle, \eta)$$
and
$$H(e_1,\dots,e_n) = \max_{e'_1 \le e_1,\ \dots,\ e'_n \le e_n}\ 
H_0(e'_1,\dots,e'_n).$$ Let us then define $r=H(2d_1-2,\dots,2d_n-2)$.
Subsection~\ref{ssec:prodEta} shows that multiplication modulo
$\langle \T \rangle$ can be performed in time
$O\tilde{~}(\delta_\d r)$. Finally, in Subsection~\ref{ssec:prec}, we
give upper bounds on $r$ that are determined by the monomial support
of~$\S$; this is the technical core of this article.

%%%%%%%%%%%%%%%%%%%%%%%%%%%%%%%%%%%%%%%%%%%%%%%%%%%%%%%%%%%%

\subsection{Computing the roots of $\S$}\label{ssec:roots}

We show here that $\S$ has all its roots in $\rng[[\eta]]$, by a
straightforward application of Hensel's lemma. 

First, we need some notation.  Given positive integers $k,\ell$ and a
subset $A \subset \rng[[\eta]]^k$, $A \bmod \eta^\ell$ denotes the set
$[a \bmod \eta^\ell \ | \ a\in A]$. Besides, we usually denote objects
over $\rng[[\eta]]$ with a ${}^\star$ superscript, to distinguish them
from their counterparts over $\rng$. Finally, we extend the
notation $\pi$ to denote the following projection
$$\begin{array}{rccc}
  \pi: & \rng[[\eta]]^n & \to & \rng[[\eta]]^{n-1} \\
  & (\alpha^\star_1,\dots,\alpha^\star_n) & \mapsto & (\alpha^\star_1,\dots,\alpha^\star_{n-1}).
\end{array}$$
Recall in what follows that $V\subset \rng^n$ is equiprojectable of multi-degree $\d$,
that its associated triangular set is $\U$, and that $\S=\eta \T + (1-\eta) \U$.
\begin{Proposition}\label{Prop:lift}
  There exists a unique set ${V^\star}$ in $\rng[[\eta]]^n$ such
  that the following holds:

\smallskip

  \begin{itemize}
  \item $V={V^\star} \bmod \eta$;

\smallskip

  \item ${V^\star}$ is equiprojectable of multi-degree $\d$;

\smallskip

  \item ${V^\star}$ supports interpolation;

\smallskip

  \item $\S$ is the triangular set associated to ${V^\star}$.
  \end{itemize} 
\end{Proposition}
\begin{pf}
We first claim that for $i \le n$ and $\alpha=(\alpha_1,\dots,\alpha_n)$ in $V$, the
partial derivative
$$\frac{\partial S_i}{\partial X_i}(\eta=0,\alpha_1,\dots,\alpha_n) = 
\frac{\partial U_i}{\partial X_i}(\alpha_1,\dots,\alpha_n) = 
\frac{\partial U_i}{\partial X_i}(\alpha_1,\dots,\alpha_i)$$ is non zero. 
Let indeed
$\alpha' = (\alpha_1,\dots,\alpha_{i-1}) \in \rng^{i-1}$.  Then, we
have by construction
$$U_i(\alpha_1,\dots,\alpha_{i-1},X_i) = \prod_{a \in v_{\alpha'}} (X_i-a),$$
so that the previous partial derivative equals
$$ \frac{\partial U_i}{\partial X_i}(\alpha_1,\dots,\alpha_{i-1},\alpha_i) = 
\prod_{a \in v_{\alpha'}, a\ne \alpha_i} (\alpha_i-a).$$ Since $V$
supports interpolation, this quantity is a product of units, so it is
a unit as well, establishing our claim.

Since the system $\S$ is triangular, its Jacobian determinant is the
product of the partial derivatives $\partial S_i/\partial X_i$. By the
previous remark, all these derivatives are units at $\alpha$, so the
Jacobian itself is a unit at $\alpha$. As a consequence, by Hensel's
lemma, for all $\alpha$ in $V$, there exists a unique $\alpha^\star$
in $\rng[[\eta]]^n$ such that $\alpha = \alpha^\star \bmod \eta$ and
$\S(\alpha^\star)=0$. We thus let ${V^\star} \subset \rng^n$ be the
set of all such $\alpha^\star$; hence $V={V^\star} \bmod \eta$ and
$\S$ vanishes at ${V^\star}$.

Next, we prove that ${V^\star}$ is equiprojectable of multi-degree
$\d$. By induction, we can assume that we have proved that
$\pi({V^\star})$ is equiprojectable of multi-degree
$(d_1,\dots,d_{n-1})$; it suffices to prove that for each
$\beta^\star$ in $\pi({V^\star})$, the fiber $V^\star_{\beta^\star}$
has cardinality $d_n$.

Let thus $\alpha^\star$ be in $V^\star_{\beta^\star}$. We prove that
for all $\gamma^\star$ in $V^\star$,
$\pi(\alpha^\star)=\pi(\gamma^\star)$ if and only if
$\pi(\alpha)=\pi(\gamma)$, with $\alpha=\alpha^\star \bmod \eta$ and
$\gamma=\gamma^\star \bmod \eta$.  To prove our claim, remark first
that if $\pi(\alpha^\star)=\pi(\gamma^\star)$ then
$\pi(\alpha)=\pi(\gamma)$, by reduction modulo $\eta$. Conversely,
suppose that $\pi(\alpha)=\pi(\gamma)$. Since the system $\S$ is
triangular, and since $\alpha^\star$ and $\gamma^\star$ are obtained
by lifting $\alpha$ and $\gamma$ using this system, we deduce that
$\pi(\alpha^\star)=\pi(\gamma^\star)$, as requested. Thus, ${V^\star}$
is equiprojectable of multi-degree $\d$.

Finally, we prove that ${V^\star}$ supports interpolation. This is
again done by induction: assume that the projection $\pi({V^\star})$
supports interpolation, let $\beta^\star$ be in $\pi({V^\star})$, and
let $a^\star$ and ${a'}^\star$ in $v^\star_\beta$. By assumption on
$V$, $a-a' \bmod \eta$ is a unit in $\rng$; thus, by Hensel's lemma,
$a^\star-{a'}^\star$ is a unit in $\rng[[\eta]]$, as requested.

This proves the existence of ${V^\star}$ with the requested
properties. Uniqueness follows in a straightforward manner from the
uniqueness property of Hensel's lemma.
\end{pf}

\noindent We continue with complexity estimates: we prove that the
roots of $\S$ can be computed in quasi-linear time.
\begin{Proposition}
  Given $\T$, $V$ and $\ell > 0$, one can compute ${V^\star} \bmod
  \eta^\ell$ in time $O(\delta_\d \L(\d)\M(\ell))$.
\end{Proposition}
\begin{pf}
  As before, we proceed inductively: we suppose that the projection
  $W^\star=\pi({V^\star})$ is known modulo $\eta^\ell$, and show how
  to deduce ${V^\star} \bmod \eta^\ell$. To do so, we evaluate all
  coefficients of $S_n$ at all points of $W^\star$ modulo
  $\eta^\ell$. Then, for each $\beta^\star$ in $W^\star$, it suffices
  to use Hensel's lemma to lift the roots of $S_n(\beta^\star,X_n)$ at
  precision $\ell$. The pseudo-code is in Figure~\ref{Fig:2_4}; for
  simplicity, we write there $W^\star$ instead of $W^\star \bmod
  \eta^\ell$.

\begin{figure}[!!!h]
\begin{center}
\fbox{
\begin{minipage}{5 cm}
\begin{tabbing}
\qquad \= \qquad \= \quad \= \quad \kill
$\underline{{\sf LiftRootsMultivariate}(V, \S, \ell)}$\\[1mm]
1 \> $n=|\S|$\\
2 \> if $n = 0$ return $[]$\\
3 \> $W^\star \leftarrow {{\sf LiftRootsMultivariate}(\pi(V), (S_1,\dots,S_{n-1}), \ell)}$ \\
4 \> for $i=0,\dots,d_n-1$ do \\
4.1 \> \> ${\rm val}_{i} \leftarrow \Eval({\rm coeff}(S_n,X_n,i),W^\star)$\\
\> \> $\slash\hspace{-1mm}*\ \text{~all computations are done modulo $\eta^\ell$~} *\hspace{-1mm}\slash$\\
5 \> for $\beta^\star$ in $W^\star$ do\\
5.1 \> \> $S_{\beta^\star} \leftarrow \sum_{i < d_n} {\rm val}_{i,\beta^\star} X_n^i + X_n^{d_n}$\\
5.2 \> \> $v^\star_{\beta^\star} \leftarrow {\sf LiftRoots}(S_{\beta^\star}, v_\beta, \ell)$\\
6 \> return $[v^\star_{\beta^\star} \ | \ \beta^\star \in W^\star]$
\end{tabbing}
\end{minipage}
}
\caption{Lifting the roots of $\S$.}
\label{Fig:2_4}
\end{center}
\end{figure}

Lemma~\ref{lemma:lift} shows that we can lift the power series roots
of a bivariate polynomial of degree $d$ at precision $\ell$ in time
$\C(d)\M(\ell)$. As a consequence, the overall cost $\C_\Lift$ of the
lifting process satisfies
$$
\begin{array}{rcl}
\C_\Lift(d_1,\dots,d_n,\ell) & ~\le~&  
\C_\Lift(d_1,\dots,d_{n-1},\ell) + 
\C_\Eval(d_1,\dots,d_{n-1}) d_n \M(\ell)\\
&&+d_1\cdots d_{n-1} \C(d_n) \M(\ell);
\end{array}
$$
the middle term gives the cost of evaluating the coefficients of $S_n$
at $W^\star \bmod \eta^\ell$ (so we apply our evaluation algorithm
with power series coefficients); and the right-hand term gives the
cost of lifting the roots of $S_n$.  This gives
$$\C_\Lift(d_1,\dots,d_n,\ell)\le 
\sum_{i \le n} \C_\Eval(d_1,\dots,d_{i-1}) d_i \M(\ell) + \sum_{i \le
  n} d_1\cdots d_{i-1} \C(d_i) \M(\ell).$$ As in the proof of
Proposition~\ref{prop:van}, one deduces that the overall sum is bounded by
$$3   \delta_\d\sum_{i \le n} \frac{\C(d_i)}{d_i}\M(\ell)
= 3   \delta_\d \L(\d)\M(\ell),$$
which concludes the proof.
\end{pf}

%%%%%%%%%%%%%%%%%%%%%%%%%%%%%%%%%%%%%%%%%%%%%%%%%%%%%%%%%%%%

\subsection{Multiplication by homotopy}\label{ssec:prodEta}

We continue with the same notation as before. To multiply two
polynomials $A,B \in \Span(M_\d)$ modulo $\langle \T \rangle$,
we may multiply them modulo $\langle \S \rangle$ over $\rng[\eta]$ and
let $\eta=1$ in the result. Now the results of the multiplication modulo
$\langle \S \rangle$ over $\rng[\eta]$ and over $\rng[[\eta]]$ are the same.
When working over $\rng[[\eta]]$, we may use the evaluation-interpolation
techniques from Subsection~\ref{ssec:mult}. Indeed,
by Proposition~\ref{Prop:lift}, $\S$ is associated to a subset
${V^\star}$ of $\rng[[\eta]]^n$ that supports interpolation.

Of course, when multiplying $A$ and $B$ modulo $\langle \S \rangle$
over $\rng[[\eta]]$, we cannot compute with (infinite) power series,
but rather with their truncations at a suitable order.  On the one
hand, this order should be larger than the largest degree of a
coefficient of the multiplication of $A$ and $B$ modulo $\langle \S
\rangle$ over $\rng[\eta]$. On the other hand, this order will
determine the cost of the multiplication algorithm, so it should be
kept to a minimum.  For $\e=(e_1,\dots,e_n)$ in $\N^n$, we define
$$ H_0(e_1,\dots,e_n) =
   \deg(X_1^{e_1}\cdots X_n^{e_n} \bmod \langle \S \rangle, \eta)$$
and
$$ H(e_1,\dots,e_n) =
   \max_{e'_1 \le e_1,\ \dots,\ e'_n \le e_n}\ H_0(e'_1,\dots,e'_n).$$ 
The following proposition relates the cost of our algorithm to
the function $H$; the behavior of this function is
studied in the next subsection.

\begin{Proposition}\label{coro:mul}
  Given $A, B$, $\T$ and $V$, one can compute $AB \bmod \langle \T
  \rangle$ in time $O(\delta_\d \L(\d)\M(r)) \subset
  O\tilde{~}(\delta_\d r)$, with $r=H(2d_1-2,\dots,2d_n-2)$.
\end{Proposition}
\begin{pf}
  The algorithm is simple: we compute $\U$ and use it to obtain
  $V^\star$ at a high enough precision. In $\rng[X_1,\dots,X_n]$, the
  product $AB$ satisfies $\deg(AB,X_i) \le 2d_i-2$ for all $i\le n$;
  since the multiplication algorithm does not perform any division by
  $\eta$, it suffices to apply it with coefficients in
  $\rng[\eta]/\langle \eta^{r+1}\rangle$, with
  $r=H(2d_1-2,\dots,2d_n-2)$. The resulting algorithm is given in
  Figure~\ref{Fig:2_5}; as before, we write $V^\star$ for simplicity,
  whereas we should write $V^\star \bmod \eta^{r+1}$.

\begin{figure}[!!!h]
\begin{center}
\fbox{
\begin{minipage}{5 cm}
\begin{tabbing}
\qquad \= \qquad \= \quad \= \quad \kill
$\underline{{\rm Mul}(A, B, \T, V)}$\\[1mm]
0. \> $\U \leftarrow {\sf AssociatedTriangularSet}(V,n)$\\
1. \> $\S \leftarrow \eta \T + (1-\eta) \U$\\
2. \> ${V^\star} \leftarrow {\sf LiftRootsMultivariate}(V, \S, r+1)$\\
3. \> $C_\eta \leftarrow {\sf Mul}(A, B, {V^\star})$\\
4. \> {\sf return} $C_\eta(1,X_1,\dots,X_n)$\\
\> $\slash\hspace{-1mm}*\ \text{$C_\eta$ is seen in $\rng[\eta][X_1,\dots,X_n]$~} *\hspace{-1mm}\slash$
\end{tabbing}
\end{minipage}
}
\caption{Multiplication algorithm.}
\label{Fig:2_5}
\end{center}
\end{figure}

The computation of $\U$ takes time $O(\delta_\d\L(\d))$ by
Proposition~\ref{prop:van}; that of $\S$ takes time
$O(\delta_\d)$. Computing $V^\star \bmod \eta^{r+1}$ takes time
$O(\delta_\d\L(\d)\M(r))$ by Proposition~\ref{Prop:lift}. Finally, the
modular multiplication takes time $O(\delta_\d\L(\d)\M(r))$ by
Proposition~\ref{prop:mul}; remark that this algorithm is run with
coefficients in $\rng[[\eta]]/\langle \eta^{r+1}\rangle$, where all
arithmetic operations take time $O(\M(r))$. Finally, specializing
$\eta$ at $1$ takes time $O(\delta_\d r)$. Summing all these costs
gives our result.
\end{pf}

%%%%%%%%%%%%%%%%%%%%%%%%%%%%%%%%%%%%%%%%%%%%%%%%%%%%%%%%%%%%

\subsection{Precision analysis}\label{ssec:prec}

We show finally how the monomial structure of the polynomials in $\S$
affects the cost of the algorithm, by means of the integer $r$ of
Proposition~\ref{coro:mul}. For $i \le n$ and
$\nu=(\nu_1,\dots,\nu_i)$ in $\N^i$, we will use the notation
$\X_i^\nu=X_1^{\nu_1} \cdots X_i^{\nu_i}$ and we write the monomial
expansion of $S_i$ as
\begin{equation}\label{eq:Si}
S_i = X_i^{d_i} + \sum_{\nu \in E_i} s_\nu \X_i^\nu,
\end{equation}
where $E_i$ is the set of exponents that appear in $S_i$, the
exponents $\nu$ are in $\N^i$, and $s_\nu$ is linear in~$\eta$.
Let us further introduce the coefficients $h_i$ defined by $h_0=0$ and
for $i \ge 1$,
\begin{equation}\label{eq:hi}
h_i = \max_{\nu \in E_i}  \frac{h_1 \nu_{1} + \cdots + h_{i-1} \nu_{i-1} + 1}{d_i-\nu_{i}}.
\end{equation}
One easily checks that all $h_i$ are positive. The following
proposition shows that through the coefficients $h_i$, the support
$E_i$ determines the cost of our algorithm.
\begin{Proposition}\label{prop:9}
 The inequality 
 $$H(e_1,\dots,e_n) \le h_1 e_1 + \cdots + h_n e_n$$
 holds for all $(e_1,\dots,e_n) \in \N^n$.
\end{Proposition}
Using Proposition~\ref{coro:mul}, this proposition gives as an easy
corollary the following statement, where we take $e_i = 2d_i -2 \le 2
d_i$; we continue using the previous notation $\T$ and $V$.
\begin{Corollary}\label{coro:r}
  Given $A, B$, $\T$ and $V$, one can compute $AB \bmod \langle \T
  \rangle$ in time $O(\delta_\d \L(\d)\M(r)) \subset
  O\tilde{~}(\delta_\d r)$, with $r \le 2( h_1 d_1 + \cdots + h_n d_n )$.
\end{Corollary}
Hence, the lower the $h_i$ the better. However, without putting extra
assumptions on the monomial supports $E_i$, Corollary~\ref{coro:r}
only yields estimates of little interest. Even in sparse cases, it
remains difficult to simplify the recurrence giving the coefficients
$h_i$. Still, several examples in the next section will show that for
some useful families of monomial supports, significantly sharper
bounds can be derived.

% We can further simplify (and weaken) the expressions:
% letting $h'_i = \max (1, h_1,\dots,h_i)$, we have
% \begin{equation}
%   \label{eq:simpl}
% h'_i\ \le\ h'_{i-1} \max_{\nu \in E'_i}  \frac{\nu_{1} + \cdots + \nu_{i-1} + 1}{-\nu_i}
% \quad\text{and~thus}\quad H(2d_1-2,\dots,2d_n-2)\ \le\ 2 d h'_n,
% \end{equation}
% with $d=\max_{i \le n} d_i$.

\medskip

The rest of this section is devoted to prove Proposition~\ref{prop:9}.
In all that follows, the multi-degree $\d=(d_1,\dots,d_n)$ and the
supports $E_i$ are fixed. We also let $E'_i$ be the set of modified
exponents
$$E'_i = \{ \nu - (0,\dots,0,d_i) \in \Z^i \ | \ \nu \in E_i\},$$
so that for all $\nu=(\nu_1,\dots,\nu_i)$ in $E'_i$, $\nu_j \ge 0$
for $j < i$ and $\nu_i < 0$.  Hence, Equation~\eqref{eq:hi} takes the
(slightly more handy) form 
\begin{equation}
h_i = \max_{\nu \in E'_i}  \frac{h_1 \nu_{1} + \cdots + h_{i-1} \nu_{i-1} + 1}{-\nu_{i}}.
\end{equation}

Recall that the function $H_0$ was defined
in the previous subsection with domain $\N^n$; in what follows, we
also consider $H_0$ as a function over $\N^i$, for $1 \le i \le n$, by
defining $H_0(e_1,\dots,e_i) = H_0(e_1,\dots,e_i,0,\dots,0)$, where
the right-hand expression contains $n-i$ zeros; for completeness, we
write $H_0()=0$ for $i=0$.  The following recurrence relation enables
us to control the growth of $H_0$.
\begin{Lemma}\label{lemma:10}
  For $i\ge 1$, let $\e=(e_1,\dots,e_i)$ be in $\N^i$ and let
  $\e'=(e_1,\dots,e_{i-1})$ in $\N^{i-1}$. Then the following
  (in)equalities hold:
$$ H_0(\e) = H_0(\e') \quad\text{if}\quad e_i < d_i,\qquad
H_0(\e) \le 1 + \max_{\nu \in E'_i}\  H_0(\e + \nu)\quad\text{otherwise.}$$
\end{Lemma}
\begin{pf}
Let us first suppose $e_i < d_i$; then, 
$$X_1^{e_1}\cdots X_i^{e_i} \bmod \langle \S \rangle= (X_1^{e_1}\cdots
X_{i-1}^{e_{i-1}} \bmod \langle \S \rangle) X_i^{e_i},$$ since the
latter product is reduced modulo $\langle \S \rangle$. Both sides have
thus the same degree in $\eta$, and our first claim follows.

We can now focus on the case $e_i \ge d_i$, for which we write $f_i =
e_i - d_i$, so that $f_i \ge 0$. From Equation~\eqref{eq:Si},
we deduce
$$X_i^{f_i} S_i =
X_i^{f_i+d_i} + \sum_{\nu \in E_i} s_\nu X_i^{f_i} \X_i^\nu,$$
and thus we get
$$X_i^{f_i} S_i = X_i^{e_i} + \sum_{\nu \in E'_i} s_\nu X_i^{e_i} \X_i^\nu,$$
by the definition of $E'_i$. In our notation, we have
$X_1^{e_1}\cdots X_{i}^{e_{i}}=\X_i^\e$. Thus, after multiplication by
$X_1^{e_1}\cdots X_{i-1}^{e_{i-1}}$ and term reorganization, the former
equality implies that
$$\X_i^\e - X_1^{e_1}\cdots X_{i-1}^{e_{i-1}} X_i^{f_i} S_i 
= -\sum_{\nu \in E'_i} s_\nu \X_i^{\e+\nu}.$$
As a consequence, we deduce that
$$\deg(\X_i^\e \bmod \langle \S \rangle, \eta)
\ \le\ \max_{\nu \in E'_i} \deg(s_\nu \X_i^{\e+\nu} \bmod \langle \S \rangle, \eta).$$
Since for $\nu$ in $E'_i$, we have
$$\deg(s_\nu \X_i^{\e+\nu} \bmod \langle \S \rangle, \eta)
=\deg(s_\nu,\eta) + \deg(\X_i^{\e+\nu} \bmod \langle \S \rangle, \eta)
=1+H_0(\e+\nu),$$
the conclusion follows. 
\end{pf}

Iterating the process of the previous lemma, we obtain the following
bound. In the next lemma, $(f_\nu)_{\nu \in E'_i}$ are a family of
integer valued variables.
\begin{Lemma}\label{lemma:3}
  Let $\e=(e_1,\dots,e_i)$ be in $\N^i$. Then the following inequality holds:
$$H_0(\e) \le \max_{\small
\begin{array}{c}
(f_\nu)_{\nu \in E'_i} \text{~non-negative integers}\\
\text{ such~that~}  0 \le e_i + \sum_{\nu \in E'_i} f_\nu \nu_i \le d_i-1
\end{array}
}\ 
 H_0(\e + \sum_{\nu \in E'_i} f_\nu \nu)+ \sum_{\nu \in E'_i} f_\nu.$$
\end{Lemma}
\begin{pf} We prove the claim by induction on $e_i$.  For $e_i \le
  d_i-1$, the family $(f_\nu=0)_{\nu \in E'_i}$ satisfies 
the constraint $0 \le e_i + \sum_{\nu \in E'_i} f_\nu \nu_i \le d_i-1$;
for this choice, the value of the function we maximize is precisely
$H_0(\e)$, so our claim holds.
Suppose now that $e_i \ge d_i$. Then, the previous 
lemma gives
\begin{equation}\label{eq:fnu}H_0(\e) \le 1 + \max_{\nu \in E'_i}\  H_0(\e + \nu).
\end{equation}
Let us fix $\nu$ in $E'_i$; then $\e+\nu$ has non-negative integer
coordinates, and its $i$th coordinate is less than $e_i$.
Thus, we can apply the induction assumption, obtaining
$$H_0(\e+\nu) \le \max_{\small
\begin{array}{c}
(f_{\nu'})_{\nu' \in E'_i} \text{~non-negative integers}\\
\text{such~that~}  0 \le e_i+\nu_i + \sum_{\nu' \in E'_i} f_{\nu'} \nu'_i \le d_i-1
\end{array}
}\ H_0(\e + \nu + \sum_{\nu' \in E'_i} f_{\nu'} \nu')+ \sum_{\nu' \in
  E'_i} f_{\nu'}.$$ To any set of non-negative integers
$(f_{\nu'})_{\nu' \in E'_i}$ with $$0 \le e_i+\nu_i + \sum_{\nu' \in
  E'_i} f_{\nu'} \nu'_i \le d_i-1$$ appearing in the previous maximum, we
associate the non-negative integers $(f'_{\nu'})_{\nu' \in E'_i}$, with
$f'_{\nu}=f_{\nu}+1$ and $f'_{\nu'}=f_{\nu'}$ otherwise. These new
integers satisfy
$$0 \le e_i + \sum_{\nu' \in  E'_i} f'_{\nu'} \nu'_i \le d_i-1$$
and
$$
H_0(\e + \nu + \sum_{\nu' \in E'_i} f_{\nu'} \nu')+ \sum_{\nu' \in
 E'_i} f_{\nu'}=
H_0(\e + \sum_{\nu' \in E'_i} f'_{\nu'} \nu')+ \sum_{\nu' \in
  E'_i} f'_{\nu'}-1.
$$
Taking maxima, we deduce from the previous inequality
$$H_0(\e+\nu) \le  \max_{\small \begin{array}{c}
(f'_{\nu'})_{\nu' \in E'_i} \text{~non-negative integers}\\
\text{such~that~}  0 \le e_i + \sum_{\nu' \in E'_i} f'_{\nu'} \nu'_i \le d_i-1
\end{array}}
\ H_0(\e + \sum_{\nu' \in E'_i} f'_{\nu'} \nu')+ \sum_{\nu' \in
  E'_i} f'_{\nu'}-1.$$
Substituting in Equation~\eqref{eq:fnu} and taking the maximum over
$\nu$ in $E'_i$ concludes the proof.
\end{pf}

For $i \le n$, let $L_i$ be the linear form $(e_1,\dots,e_i) \mapsto
h_1 e_1 + \cdots + h_{i} e_{i}$, where the $h_i$ are as in
Equation~\eqref{eq:hi}. The following lemma concludes the proof of
Proposition~\ref{prop:9}; as we did for $H_0$, for $i\le n$, we extend
$H$ to $\N^i$, by writing
$H(e_1,\dots,e_i)=H(e_1,\dots,e_i,0,\dots,0)$.
\begin{Lemma}
  For $i \le n$ and $\e=(e_1,\dots,e_i)$ in $\N^i$, the inequality
  $H(\e)\le L_i(\e)$ holds.
\end{Lemma}
\begin{pf}
  It is sufficient to prove that $H_0(\e)\le L_i(\e)$ holds; since all
  coefficients of $L_i$ are non-negative, $L_i$ is non-decreasing with
  respect to all of its variables, which implies the thesis.  

  We prove our inequalities by induction on $i\ge 0$. For $i=0$, we
  have $H_0()=L_0()=0$; hence, our claim vacuously holds at this
  index.
%  For
% $i=1$, so that $\e=(e_1)$, we are to prove that 
% $$H_0(e_1) = \deg(X_1^{e_1} \bmod S_1,\eta)$$
% admits the upper bound $h_1 e_1$, where $h_1$ is defined as
% $$
% S_1 = X_1^{d_1} + \sum_{\nu_1 \in E_1} X_1^{\nu_1}
% \quad\text{and}\quad
% h_1 = \max_{\nu_1 \in E'_1}  \frac{1}{-\nu_1}
% = \max_{\nu_1 \in E_1}  \frac{1}{d_1-\nu_1}.
% $$
% Now, for $e_1 < d_1$, $\deg(X_1^{e_1} \bmod S_1,\eta)=0$.  For $e_1
% \ge d_1$, Lemma~\ref{lemma:10} shows that
% $$H_0(e_1)=1 + \max_{\nu_1 \in E_1} H_0(e_1+\nu_1-d_1)=1+H_0(e_1+1/h_1).$$
  For $i \ge 1$, we now prove that if our inequality holds at index
  $i-1$, it will also hold at index $i$. Lemma~\ref{lemma:3} shows
  that for any ${\e} \in \N^i$, we have the inequality
$$H_0(\e) \le \max_{\small
\begin{array}{c}
(f_\nu)_{\nu \in E'_i} \text{~non-negative integers}\\
\text{such~that~}  0 \le e_i + \sum_{\nu \in E'_i} f_\nu \nu_i \le d_i-1
\end{array}
}\ H_0(\e + \sum_{\nu \in E'_i} f_\nu \nu)+ \sum_{\nu \in E'_i}
f_\nu.$$ Let $\varphi$ be the natural projection $\N^i \to \N^{i-1}$,
let $(f_\nu)_{\nu \in E'_i}$ be non-negative integers that satisfy the
conditions in the previous inequality. Since $\e + \sum_{\nu \in E'_i}
f_\nu \nu$ has degree in $X_i$ less than $d_i$, the first point of
Lemma~\ref{lemma:10} shows that
$$H_0(\e + \sum_{\nu \in E'_i} f_\nu \nu) \ = \ 
H_0(\varphi(\e + \sum_{\nu \in E'_i} f_\nu \nu));$$ 
the induction assumption implies that this quantity 
is bounded from above by 
$$L_{i-1}(\varphi(\e + \sum_{\nu \in E'_i} f_\nu \nu)).$$
As a consequence, $H_0(\e)$ admits the upper
bound 
$$\max_{\small
\begin{array}{c}
(f_\nu)_{\nu \in E'_i} \text{~non-negative integers}\\
\text{such~that~}  0 \le e_i + \sum_{\nu \in E'_i} f_\nu \nu_i \le d_i-1
\end{array}
}\ L_{i-1}(\varphi(\e + \sum_{\nu \in E'_i} f_\nu \nu))+\sum_{\nu \in
  E'_i} f_\nu.$$ This quantity itself is upper-bounded by a similar
expression, where we allow the $f_\nu$ to be non-negative reals numbers;
this gives
$$H_0(\e) \ \le \ \max_{\small
\begin{array}{c}
(f_\nu)_{\nu \in E'_i} \text{~non-negative real numbers} \\ 
\text{such~that~} 0 \le e_i + \sum_{\nu \in E'_i} f_\nu \nu_i \le d_i-1
\end{array}
}\ L_{i-1}(\varphi(\e + \sum_{\nu \in E'_i} f_\nu \nu))+\sum_{\nu \in
  E'_i} f_\nu.$$ Since all $h_i$ and all $\nu_1,\dots,\nu_{i-1}$ are
non-negative, the function of $(f_\nu)_{\nu \in E'_i}$ we want to
maximize is affine with non-negative coefficients. The domain where we
maximize it is defined by the conditions
$$f_\nu \ge 0 \text{~for all~} \nu \in E'_i, \quad 0 \le  e_i + \sum_{\nu \in E'_i}  f_\nu \nu_i \le d_i-1,$$
and it is contained in the domain $D$ defined by the conditions
$$f_\nu \ge 0 \text{~for all~} \nu \in E'_i, \quad 0 \le e_i + \sum_{\nu \in E'_i}  f_\nu \nu_i.$$
Since all unknowns $f_\nu$ are non-negative, while the coefficients
$\nu_i$ are negative, the domain $D$ is convex and bounded. Hence, the
maximal value we look for is upper-bounded by the maximal value at the
end-vertices of $D$, distinct from the origin; these vertices are 
$$E_\nu = \{f_{\nu'} = 0 \text{~~for~~} \nu'\ne \nu,\qquad f_\nu = -\frac{e_i}{\nu_i}\},
\quad\text{for}\quad \nu\in E'_i.$$ At the point $E_\nu$, the
objective function  takes the value
$$ L_{i-1}(\varphi(\e - \frac{e_i}{\nu_i} \nu))  -\frac{e_i}{\nu_i}.$$
By the linearity of $L_{i-1}$ and $\varphi$, this can be rewritten as
$$  L_{i-1}(\varphi(\e)) - L_{i-1}(\varphi(\frac{e_i}{\nu_i} \nu))-\frac{e_i}{\nu_i}\ =\ 
L_{i-1}(\varphi(\e)) - \frac{L_{i-1}(\varphi(\nu))+1}{\nu_i} e_i.$$
As a consequence, we obtain the upper bound
$$H_0(\e) \ \le L_{i-1}(\varphi(\e)) + \max_{\nu \in E'_i}  \frac{L_{i-1}(\varphi(\nu))+1}{-\nu_i} e_i.$$
To simplify this further, note that the term $L_{i-1}(\varphi(\e))$
rewrites as $h_1 e_1 + \cdots + h_{i-1} e_{i-1}$.  Similarly,
$L_{i-1}(\varphi(\nu))+1$ equals $h_1 \nu_1 + \cdots + h_{i-1} \nu_{i-1}$.
We deduce the inequality
$$H_0(\e) \ \le\ h_1 e_1 + \cdots + h_{i-1} e_{i-1} \ + \
\max_{\nu \in E'_i}  \frac{h_1 \nu_1 + \cdots + h_{i-1} \nu_{i-1}+1}{-\nu_i} e_i,
$$
which we can finally rewrite as 
$$H_0(\e) \ \le\ h_1 e_1 + \cdots + h_{i-1} e_{i-1} + h_i e_i,$$
as requested.
\end{pf}

%%%%%%%%%%%%%%%%%%%%%%%%%%%%%%%%%%%%%%%%%%%%%%%%%%%%%%%%%%%%
%%%%%%%%%%%%%%%%%%%%%%%%%%%%%%%%%%%%%%%%%%%%%%%%%%%%%%%%%%%%
%%%%%%%%%%%%%%%%%%%%%%%%%%%%%%%%%%%%%%%%%%%%%%%%%%%%%%%%%%%%

\section{Examples}\label{sec:appli}

%%%%%%%%%%%%%%%%%%%%%%%%%%%%%%%%%%%%%%%%%%%%%%%%%%%%%%%%%%%%

\subsection{Main family of examples}\label{ssec:main}

We give explicit estimates for the coefficients $h_i$ of the previous
section on the following family of examples. We consider triangular
sets $\T=(T_1,\dots,T_n)$ such that $T_i$ has the form
\begin{equation}\label{eq:ex2}
  T_i = X_i^{d_i} + \sum_{\nu \in D_i} t_\nu \X_i^\nu, \quad t_\nu \in \rng,
\end{equation}
where all $\X_i^\nu$ are monomials in $X_1,\dots,X_i$ of total degree
at most $\lambda_i$, for some $\lambda_i \in \N$. We let $d=\max_{i
  \le n} d_i$, and we suppose that $\rng$ contains at least $d$
pairwise distinct values $x_1,\dots,x_d$, with $x_i-x_j$ a unit for $i
\ne j$.

The following proposition illustrates three different situations. The
first two cases display a cost quasi-linear in $d\delta_\d$, which is
satisfying, especially for small $d$; the last one shows that small
changes in the assumptions can induce large overheads.  We will see in
the next subsection cases where $d_i$ is constant equal to $d$, or
$d_i=n+1-i$; in such cases, $d$ is logarithmic in $\delta_\d$ and the
cost $O\tilde{~}(d \delta_\d)$ is thus $O\tilde{~}(\delta_\d)$, which
is what we were aiming at.

\begin{Proposition}\label{prop:family}
  With assumptions as above, multiplication modulo $\langle \T\rangle$
  can be performed with the following complexities:
$$
\begin{array}{rcll}
O\big(\,n\, \delta_\d\, \frac{\C(d)}d\, \M(nd)\,\big )\ & \subset&\ O\tilde{~}(d \delta_\d) & \text{~if $\lambda_i = d_i-1$ for all $i$},\\[1mm]
O\big(\,n\, \delta_\d\,\frac{\C(d)}d\, \M(n^2d)\,\big ) \ &\subset&\ O\tilde{~}(d \delta_\d) & \text{~if $\lambda_i = d_i$ for all $i$}, \\[1mm]
O\big(\,n\, \delta_\d\, \frac{\C(d)}d\,\M(2^nd)\,\big ) \ &\subset&\ O\tilde{~}(2^n d \delta_\d) & \text{~if $\lambda_i = d_i+1$ for all $i$}.
\end{array}
$$
\end{Proposition}
\begin{pf}
First, we construct $V$: we simply choose the grid
\begin{equation}\label{eq:V}
V = [x_1,\dots,x_{d_1}] \times \cdots \times [x_1,\dots,x_{d_n}].
\end{equation}
Thus, we have $U_i = (X_i-x_1) \cdots (X_i-x_{d_i})$; as before we let
$\S =\eta \T + (1-\eta) \U$. Thus, the monomial support $E_i$ associated with
$S_i$ is contained in
$$D'_i = D_i \cup \{(0,\dots,0,\nu_i) \ | \ 0 \le \nu_i < d_i\}.$$
Since each monomial in $D_i$ has an exponent of the form
$(\nu_1,\dots,\nu_i)$, with $\nu_1+\cdots + \nu_i \le \lambda_i$ and
$\nu_i < d_i$, we deduce from Equation~\eqref{eq:hi} that
$$h_i\ \le \
 \max_{\nu \in D'_i} \frac{ h_1 \nu_1+\cdots + h_{i-1} \nu_{i-1} + 1} {d_i-\nu_i}
\ \le \
 \max\, \Big(\max_{\nu \in D_i} \frac{ h_1 \nu_1+\cdots + h_{i-1} \nu_{i-1} + 1} {d_i-\nu_i}, \ 1\Big).$$
Let $h'_i = \max(h_1,\dots,h_i)$, so that 
\begin{equation}
  \label{eq:hprime}
h_i\ \le\ 
 \max\, \Big( \max_{\nu \in D_i} \frac{  h'_{i-1}(\nu_1+\cdots + \nu_{i-1}) + 1} {d_i-\nu_i}, 1 \Big)
\ \le\
 \max\, \Big( \max_{\nu \in D_i} \frac{  h'_{i-1}(\lambda_i  - \nu_i) + 1} {d_i-\nu_i}, 1 \Big).
\end{equation}
Knowing the distribution of the $d_i$ and $\lambda_i$, the former
relation makes it possible to analyze the growth of the coefficients
$h_i$, and thus of $ 2(d_1 h_1 + \cdots + d_n h_n)$.

\smallskip

\begin{description}
\item [Case $1$.] Suppose first that $\lambda_i = d_i-1$. Then, the
  former inequality implies $h_i \le 1$ for all $i$, so that $2(d_1
  h_1 + \cdots + d_n h_n) \le 2 n d. $

\smallskip

\item [Case $2$.] If $\lambda_i = d_i$, then~\eqref{eq:hprime} becomes
  $h_i \le h'_{i-1} + 1,$ so that $h_i \le i$ for all $i$, and thus
$2(d_1 h_1 + \cdots + d_n h_n) \le  n(n+1) d. $

\smallskip

\item [Case $3$.] If finally $\lambda_i=d_i+1$, then ~\eqref{eq:hprime}
  becomes $h_i \le 2h'_{i-1} + 1,$ so that $h'_i \le 2^i-1$. In this
  case, we get $2(d_1 h_1 + \cdots + d_n h_n) \le 2^{n+2}d. $
\end{description}

\smallskip
\noindent
To conclude the proof, we simply plug the previous estimates in the
cost estimate $O(\delta_\d \L(\d)\M(r))$ of Corollary~\ref{coro:r},
with $r \le 2(d_1 h_1 + \cdots + d_n h_n)$, and we use the upper bound
$\L(\d) \le n\C(d)/d$ of Equation~\eqref{eq:L}.
\end{pf}

%%%%%%%%%%%%%%%%%%%%%%%%%%%%%%%%%%%%%%%%%%%%%%%%%%%%%%%%%%%%

\subsection{Cauchy modules}

Cauchy modules~\cite{ReVa99} are a basic construction in Galois theory
and invariant
theory~\cite{Sturmfels93,ReVa99,AbOrReVa04,ReYo06}. Starting from a
monic polynomial $F\in \rng[X]$ of degree $d$, we define a triangular
set $F_1,\dots,F_d$ by letting $F_1(X_1) = F(X_1)$ and taking iterated
divided differences:
$$ F_{i+1}(X_1,\dots,X_{i+1})
=
\frac{F_{i}(X_1,\dots,X_{i-1},X_i)-F_{i}(X_1,\dots,X_{i-1},X_{i+1})}{X_i-X_{i+1}}
\qquad 1 \le i < d.$$ The polynomials $F_1,\dots,F_d$ form a
triangular set of multi-degree $\d=(d,d-1,\dots,1)$, so that
$\delta_\d=d!$; their interest stems from the fact that they form a
system of generators of the ideal
$(\sigma_i - (-1)^i f_{d-i})_{1 \le i\le d},$
where $\sigma_i$ is the $i$th elementary symmetric polynomial
in $X_1,\dots,X_d$ and $f_i$ is the coefficient of $X^i$ in $F$.

One easily checks that $F_i$ has total degree at most $d+1-i$.  Hence,
assuming that $0,\dots,d-1$ are units in $\rng$, we are under the
assumptions of Subsection~\ref{ssec:main}, with $\lambda_i=d_i=d+1-i$
for all $i$ and $(x_1,\dots,x_d)=(0,\dots,d-1)$. As a consequence,
Proposition~\ref{prop:family} shows that multiplication modulo
$\langle F_1,\dots,F_d \rangle$ can be done using $O\big(d!\, \C(d)
\,\M(d^3)\big )$ operations in $\rng$, that is, in quasi-linear time
$O\tilde{~}(d!)$. This improves for instance the results given
in~\cite{GaScTh06} on the evaluation properties of symmetric
polynomials.

%%%%%%%%%%%%%%%%%%%%%%%%%%%%%%%%%%%%%%%%%%%%%%%%%%%%%%%%%%%%

\subsection{Polynomial multiplication}

We show now how to derive quasi-linear time algorithms for {\em
  univariate} multiplication in $\rng[X]$ from our previous
multivariate construction. Unfortunately, our algorithm does not
improve on the complexity of Cantor-Kaltofen's
algorithm~\cite{CaKa91}; however, we believe it is worth mentioning.
Precisely, given $n \ge 1$, we give here an algorithm to perform
truncated multiplication in $\rng[X]/\langle X^{2^n}\rangle$. We
introduce variables $X_1,\dots,X_n$; computing in $A=\rng[X]/\langle
X^{2^n} \rangle$ is equivalent to computing in
$B=\rng[X_1,\dots,X_n]/\langle V_1,\dots,V_n \rangle$, with
$\V=(V_1,\dots,V_n)$ given by
$$\left | 
\begin{array}{l}
X_1-X_n^{2^{n-1}}\\
~~~\vdots\\
X_{n-1}-X_n^2\\
X_n^{2^n},
\end{array}\right .
$$
since the dummy variables $X_1,\dots,X_{n-1}$ play no role in this
representation. However, changing the order of the variables, we see
that the ideal $\langle V_1,\dots,V_n \rangle$ is also equal to the ideal
$\langle T_1,\dots,T_n \rangle$ given by
$$\left | 
\begin{array}{l}X_n^2-X_{n-1}\\
~~~\vdots\\
X_2^2-X_1\\
X_1^2.
\end{array}\right .
$$
The $\rng$-basis of $B$ corresponding to $\V$ is $(X_n^i)_{i < 2^n}$;
the basis corresponding to $\T$ is $M_\d$ (notation defined in the
introduction), with $\d=(2,\dots,2)$. Besides, the change of basis
does not use any arithmetic operation, since it amounts to rewrite
the exponents $i$ in base 2, and conversely.

Hence, we can apply our multivariate multiplication algorithm modulo
$\langle \T \rangle$. Remark that the triangular set $\T$ satisfies
the assumptions of Subsection~\ref{ssec:main} (for any $\rng$), with
$d_1=\cdots=d_n=d=2$, $\delta_\d=2^n$, $\lambda_1=\dots=\lambda_n=1$
and $(x_1,x_2)=(0,1)$.  By Proposition~\ref{prop:family}, we deduce
that the cost of a multiplication in $B$, and thus in $A$, is $O(2^n n
\M(n)).$ Since one can multiply univariate polynomials of degree $2^n$
using two multiplications in $A$, this gives the recurrence
$$\M(2^n)\ \le\ {\sf k}  2^n n \M(n) \quad\text{and thus}\quad
\M(d)\ \le\ {\sf k}'  d \log(d) \M(\log(d))
$$
for some constants ${\sf k}$, ${\sf k}'$. Unrolling the recursion 1,
2, \dots, times, and taking $\M(n)\in O(n^2)$ to end the recursion, we
obtain quasi-linear estimates of the form
$$\M(d) \in O(d\log(d)^3) \quad\text{or} \quad \M(d) \in O(d\log(d)^2\log(\log(d))^3), \quad \dots$$
The main noteworthy feature of this multiplication algorithm is that
no root of unity is present, though our multivariate
evaluation-interpolation routine is somewhat similar to a multivariate
Fourier Transform.  In particular, the case when $2$ is a zero-divisor
in $\rng$ requires no special treatment, contrary to~\cite{CaKa91}.

%%%%%%%%%%%%%%%%%%%%%%%%%%%%%%%%%%%%%%%%%%%%%%%%%%%%%%%%%%%%

\subsection{Exponential generating series multiplication}\label{ssec:expo}

We continue with a question somehow similar to the one in the previous
subsection. Given two sequences $a_0,\dots,a_d$ and $b_0,\dots,b_d$ in
$\rng$, we want to compute the sequence $c_0,\dots,c_d$ such that
\begin{equation}
  \label{eq:power0}
c_k =\sum_{i+j=k} {k \choose i} a_i b_{j},
\end{equation}
where the binomial coefficients are the coefficients of the expansion
of $(1+X)^i$ in $\rng[X]$. We discuss an application of this question
in the next section.

The naive algorithm has cost $O(d^2)$. If $1,\dots,d$ are units in
$\rng$, the former equation takes the form
\begin{equation}
  \label{eq:power}
 \sum_{i \le d} \frac{c_i}{i!}X^i\ =\ \sum_{i\le d} \frac{a_i}{i!}X^i\, \sum_{i \le d} \frac{b_i}{i!}X^i \mod X^{d+1},
\end{equation}
so we can achieve a cost $O(\M(d))$. Under some much milder
assumptions on $\rng$, we are going to see how to achieve a similar
cost through multivariate computations.

We will suppose that there exists a prime $p$ such that for $a \in
\N$, if $\gcd(a,p)=1$, then $a$ is a unit in $\rng$ (this is the case
e.g. for $\rng=\Z/p^k\Z$). Let $n$ be such that $d+1 \le p^n$, and
introduce the triangular set $\T=(T_1,\dots,T_n)$ defined by
$$\left | 
\begin{array}{l}
X_n^p-pX_{n-1}\\
~~~\vdots\\
X_2^p-pX_1\\
X_1^p.
\end{array}\right .
$$
In what follows, for $i \ge 0$, $(i_0,i_1,\dots)$ denotes the sequence
of its coefficients in base $p$; thus, for $i\le d$, only
$i_0,\dots,i_{n-1}$ can be non-zero.  Besides, we let $f:\N \to \N$ be
defined by $f(i)=i!/p^{v(i!)}$, where $v(i!)$ is the $p$-adic
valuation of $i!$. In particular, $f(i)$ is a unit in~$\rng$.
\begin{Proposition}
  Let 
$$A=\sum_{i \le d} \frac{a_i}{f(i)} X_n^{i_0} \cdots X_1^{i_{n-1}},\quad
B=\sum_{i \le d} \frac{b_i}{f(i)} X_n^{i_0} \cdots X_1^{i_{n-1}},\quad
C=\sum_{i \le d} \frac{c_i}{f(i)} X_n^{i_0} \cdots X_1^{i_{n-1}}.$$ Then
$C=AB \bmod \langle \T \rangle$.
\end{Proposition}
\begin{pf}
  Let $i,j \le d$, with $k=i+j \le d$. We start by the obvious remark that
  \begin{equation}
    \label{eq:val}
 {k \choose i} = \frac{f(k)}{f(i)f(j)}\, p^{v({k \choose i})}    
  \end{equation}
holds in $\rng$. Besides, the normal form of the product
$\frac{a_i}{f(i)} X_n^{i_0} \cdots X_1^{i_{n-1}}$ by
$\frac{b_j}{f(j)} X_n^{j_0} \cdots X_1^{j_{n-1}}$ modulo $\langle \T \rangle$ is
$$\frac{a_i b_j}{f(i)f(j)} p^{c(i,j)} X_n^{k_0} \cdots X_1^{k_{n-1}},$$
where $c_{i,j}$ is the number of carries held in the addition of $i$ and $j$ 
in base $p$. From~\cite[Eq.~(1.6)]{StMoAm08}, $c_{i,j}$ is exactly 
the valuation of the binomial coefficient $k \choose i$.
Thus, by~\eqref{eq:val}, the former product equals
$${k \choose i} \frac{a_i b_j}{f(k)}X_n^{k_0} \cdots X_1^{k_{n-1}}.$$
Summing over all $i,j$ gives our claim.
\end{pf}

As in the previous subsection, we can apply our multivariate
multiplication algorithm modulo $\langle \T \rangle$. Remark that the
triangular set $\T$ satisfies the assumptions of
Subsection~\ref{ssec:main}, with $d_1=\cdots=d_n=d=p$,
$\delta_\d=p^n$, $\lambda_1=\dots=\lambda_n=1$ and
$(x_1,\dots,x_p)=(0,\dots,p-1)$. Note as well that we can take $n\in
O(\log_p(d))$, and that $\delta_\d=p^n \le pd$.

By Proposition~\ref{prop:family}, we deduce that the cost of computing
$C$, and thus all $c_0,\dots,c_d$, is $O( d\log(d)\, \M(p)\,
\M(p\log_p(d) )).$ If $p$ is fixed, we obtain the estimate $O(d
\log(d) \M(\log(d))).$ This is not as good as the estimate
$O(\M(d))=O(d \log(d) \log\log(d))$ we obtained in characteristic
zero, but quite close.

%%%%%%%%%%%%%%%%%%%%%%%%%%%%%%%%%%%%%%%%%%%%%%%%%%%%%%%%%%%%
%%%%%%%%%%%%%%%%%%%%%%%%%%%%%%%%%%%%%%%%%%%%%%%%%%%%%%%%%%%%
%%%%%%%%%%%%%%%%%%%%%%%%%%%%%%%%%%%%%%%%%%%%%%%%%%%%%%%%%%%%

\section{Application: computing with algebraic numbers}\label{sec:algnum}

We finally present an application of the previous constructions to
computation with algebraic numbers, and give timings of our
implementation.

%%%%%%%%%%%%%%%%%%%%%%%%%%%%%%%%%%%%%%%%%%%%%%%%%%%%%%%%%%%%

\subsection{Presentation of the problem}

Let $k$ be a field and let $f$ and $g$ be monic polynomials in $k[T]$,
of degrees $m$ and $n$ respectively. We are interested in computing
their {\it composed sum} $h=f\oplus g$. This is the polynomial of
degree $d=mn$ defined by $$ f\oplus g = \prod _{\alpha, \beta}
(T-\alpha-\beta),$$ the product running over all the roots $\alpha$ of
$f$ and $\beta$ of $g$, counted with multiplicities, in an algebraic
closure $\overline{k}$ of $k$.

A natural approach consists in computing $h(T)$ as the resultant of
$f(T-U)$ and $g(U)$ in~$U$. However, the fastest algorithm for
resultants~\cite{Reischert97} has a complexity of
order~$O\tilde{~}(d^{1.5})$ for $m=n$. To do better, Dvornicich and
Traverso~\cite{DvTr87} suggested to compute the power sums
$$a_i=\sum_{f(\alpha)=0} \alpha^i,\quad b_i=\sum_{g(\beta)=0} \beta^i$$ of
respectively $f$ and $g$, and deduce the power sums $c_i$ of $h$, by
means of Equation~\eqref{eq:power0}.
In~\cite{BoFlSaSc06}, this approach is showed to take time $O(\M(d))$,
over fields of characteristic zero or larger than~$d$.  Indeed,
computing $(a_i)_{i \le d}$ and $(b_i)_{i \le d}$ can be done in
$O(\M(d))$ operations, over any field, using Newton iteration for
power series division~\cite{Schonhage82}. Then, by our assumption on
the characteristic, one can compute $(c_i)_{i \le d}$ in quasi-linear
time using Equation~\eqref{eq:power}, for another $\M(d)+O(d)$
operations. Finally, knowing $(c_i)_{i \le d}$, one can then recover
$h$ in time $O(\M(d))$ as well, using fast exponential
computation~\cite{Brent75,Schonhage82,vdH:fnewton,BS:exp};
this step relies as well on the assumption on the characteristic.

If $k$ has positive characteristic less than $d$, two issues arise:
Equation~\eqref{eq:power} makes no sense anymore and $(c_i)_{i \le d}$
are actually not enough to recover~$h$. To our knowledge, no general
solution better than the resultant method was known up to now (partial
answers are in~\cite{BoFlSaSc06,Schost05} under restrictive
conditions). We propose here a solution that works over finite fields,
following an idea introduced in~\cite{GoPe04}.

For simplicity, we consider only $k=\F_p$. Since our algorithm
actually does computations over rings of the form $\Z/p^\alpha \Z$,
measuring its complexity in $\F_p$-operations as we did up to now is
not appropriate. Instead, we count bit operations. Thus, we let
$\M_\Z$ be such that integers of bit-length $\ell$ can be multiplied
using $\M_\Z(\ell)$ bit operations; quasi-linear estimates are known
as well for $\M_\Z$, the best to date being F{\"u}rer's $\ell
\log(\ell)2^{O(\log^*(\ell))}$~\cite{Furer07}.

\begin{Proposition}
  Given $f$ and $g$, one can compute $h$ using
$$O\big( (\M(d) \, + \,  d\, \log(d)\, \M(p)\, \M(p\log_d(p))) \ {\sf N}(p,d)\big)$$
  bit operations, with ${\sf N}(p,d)=O(\M_\Z(\log(p))\log(\log(p)) + \M_\Z(\log(d)))$.
\end{Proposition}
\noindent After simplification, this cost is seen to be
$O\tilde{~}(dp^2)$ bit operations. Also, if we consider $p$ fixed, the cost
becomes
$$O\big(\,(\M(d)\,+\, d \log(d)\M(\log(d)))\,\M_\Z(\log(d))\,\big),$$
that is, quasi-linear.

\begin{pf}
  Let $\Z_p$ be the ring of $p$-adic integers and let $F$ and $G$ be
  monic lifts of $f$ and $g$ in $\Z_p[T]$, of degrees $m$ and
  $n$. Defining $H=F \oplus G \in \Z_p[T]$, we have that $h=H \bmod
  p$. Let further $(A_i)_{i \ge 0}$, $(B_i)_{i \ge 0}$ and $(C_i)_{i
    \ge 0}$ be the power sums of respectively $F$, $G$ and $H$. For
  any $\alpha \ge 0$, the reductions $A_i \bmod p^\alpha$, $B_i \bmod
  p^\alpha$, and $C_i \bmod p^\alpha$ satisfy
  Equation~\eqref{eq:power0}, so we can apply the results of
  Subsection~\ref{ssec:expo} to deduce $(C_i \bmod p^\alpha)_{i \le
    d}$ from $(A_i \bmod p^\alpha)_{i\le d}$ and $(B_i \bmod
  p^\alpha)_{i \le d}$.

  Besides, taking $\alpha= \lfloor\log_p (d) \rfloor+1$, it is proved
  in~\cite{BoGoPeSc05} that given $(C_i \bmod p^\alpha)_{i \le d}$,
  one can compute $h$ in quasi-linear time $O(\M(d) \M_\Z(\log_p(d)))$
  bit operations. Remark that this step is non trivial: recovering a
  polynomial of degree $d$ from its Newton sums requires divisions by
  $1,\dots,d$, and not all these numbers are units in small
  characteristic.

  In the algorithm, the function {\sf Lift} simply lifts its argument
  from $\F_p[T]=\Z/p\Z[T]$ to $\Z/p^\alpha\Z[T]$; the function {\sf
    PowerSums} computes the first $d$ power sums of its arguments by
  the algorithm of~\cite{Schonhage82}. Step 7 applies the algorithm of
  Subsection~\ref{ssec:expo}, and the last step uses the algorithm
  presented in~\cite{BoGoPeSc05} to recover $h$.

  Our choice of $\alpha$ implies that $\log(p^\alpha)=O(\log(d))$.
  Thus, operations $(+,\times)$ modulo $p^\alpha$ take
  $O(\M_\Z(\log(d)))$ bit operations~\cite[Chapter~9]{GaGe99}.  Using
  Newton iteration, inversions modulo $p^\alpha$ take ${\sf
    N}(p,d)=O(\M_\Z(\log(p))\log(\log(p)) + \M_\Z(\log(d)))$ bit
  operations, where the first term stands for the cost computing the
  inverse modulo $p$, and the second one for lifting it modulo
  $p^\alpha$.

  The cost of computing $(A_i)_{i \le d}$ and $(B_i)_{i \le d}$ is
  $O(\M(d))$ operations modulo $p^\alpha$; this dominates the cost of
  recovering $h$. The remaining cost is that of computing $(C_i)_{i
    \le d}$, which is reported in Subsection~\ref{ssec:expo} in terms
  of numbers of operations modulo $p^\alpha$. The previous estimate on
  ${\sf N}(p,d)$ concludes the proof.
\end{pf}

\begin{figure}[!!!h]
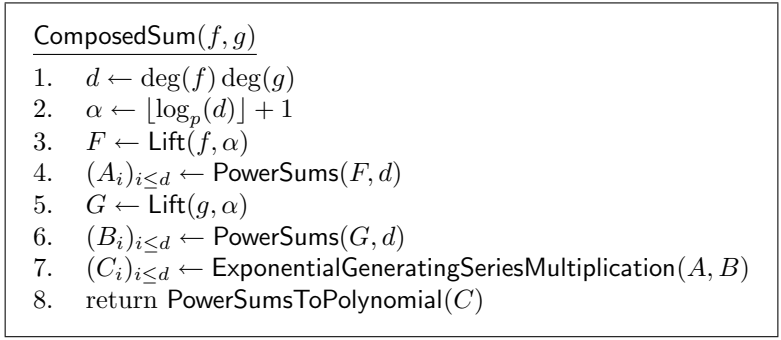

\begin{center}
\fbox{
\begin{minipage}{8 cm}
\begin{tabbing}
\qquad \= \qquad \= \quad \= \quad \kill
$\underline{{\sf ComposedSum}(f,g)}$\\[1mm]
1. \> $d \leftarrow \deg(f)\deg(g)$\\
2. \> $\alpha \leftarrow \lfloor\log_p (d) \rfloor+1$\\
3. \> $F \leftarrow {\sf Lift}(f, \alpha)$\\
4. \> $(A_i)_{i \le d} \leftarrow {\sf PowerSums}(F, d)$\\
5. \> $G \leftarrow {\sf Lift}(g, \alpha)$\\
6. \> $(B_i)_{i \le d} \leftarrow {\sf PowerSums}(G, d)$\\
7. \> $(C_i)_{i \le d} \leftarrow {\sf ExponentialGeneratingSeriesMultiplication}(A,B)$\\
8. \> return ${\sf PowerSumsToPolynomial}(C)$
\end{tabbing}
\end{minipage}
}
\caption{Composed sum in small characteristic.}
\label{Fig:composed_sum}
\end{center}
\end{figure}

%%%%%%%%%%%%%%%%%%%%%%%%%%%%%%%%%%%%%%%%%%%%%%%%%%%%%%%%%%%%

\subsection{Experimental results}

We implemented the composed sum algorithm over $\F_2$ ({\it i.e.},
$p=2$ here). We used the NTL C++ package as a basis~\cite{NTL}. Since
NTL does not implement bivariate resultants, we also used
Magma~\cite{magma} for comparison with the resultant method. All
timings are obtained on an AMD Athlon 64 with 5GB of RAM.

\begin{figure}[!!!h]
\begin{center}
\ifpdf
\includegraphics[width=15cm]{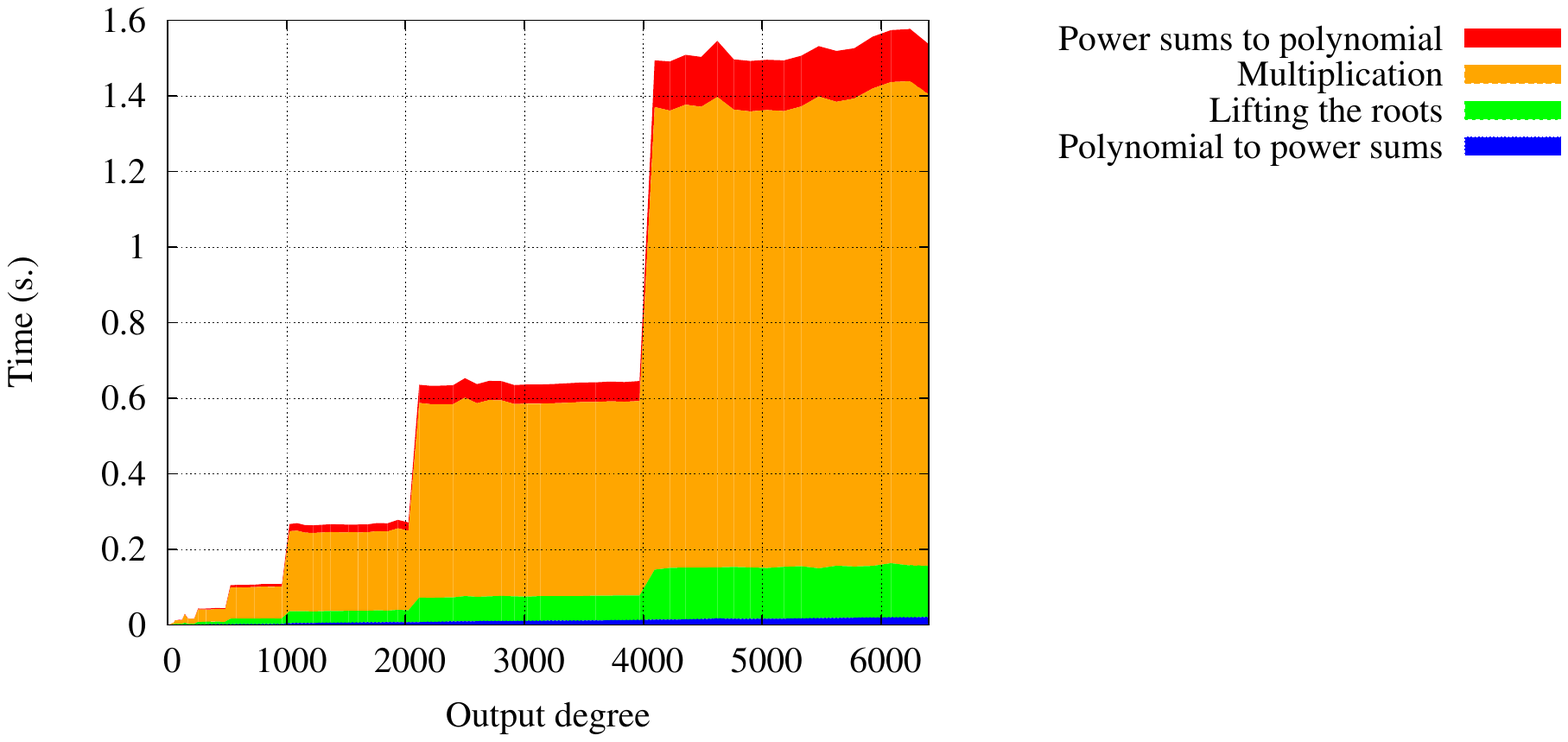}  
\else
\includegraphics[width=15cm]{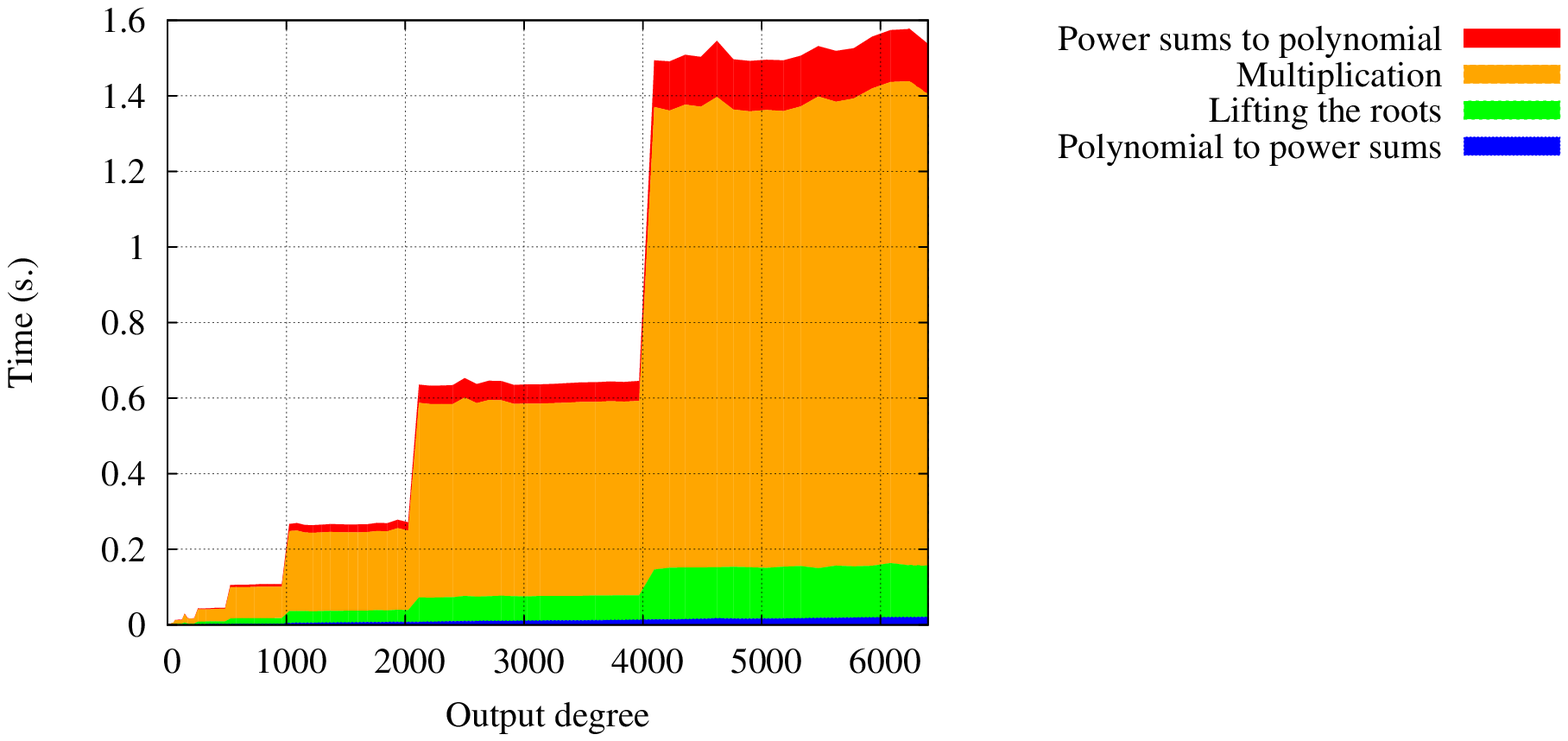}  
\fi
\end{center}
\caption{Detailed timings for our algorithm}
\label{fig:detailed}
\end{figure}

Figure~\ref{fig:detailed} gives detailed timings for our algorithm;
each colored area gives the time of one of the main tasks. The less
costly step is the first, the conversion from the original polynomials
to their Newton sums. Then, we give the time needed to compute all the
power series roots needed for our multiplication algorithm, followed
by the evaluation-interpolation process itself; finally, we give the
time necessary to recover $h$ from its power sums. Altogether, the
practical behavior of our algorithm matches the quasi-linear
complexity estimates. The steps we observe correspond to the increase
in the number of variables in our multivariate polynomials, and are
the analogues of the steps observed in classical FFT.

\begin{figure}[!!!h]
\begin{center}
\ifpdf
\includegraphics[width=9cm]{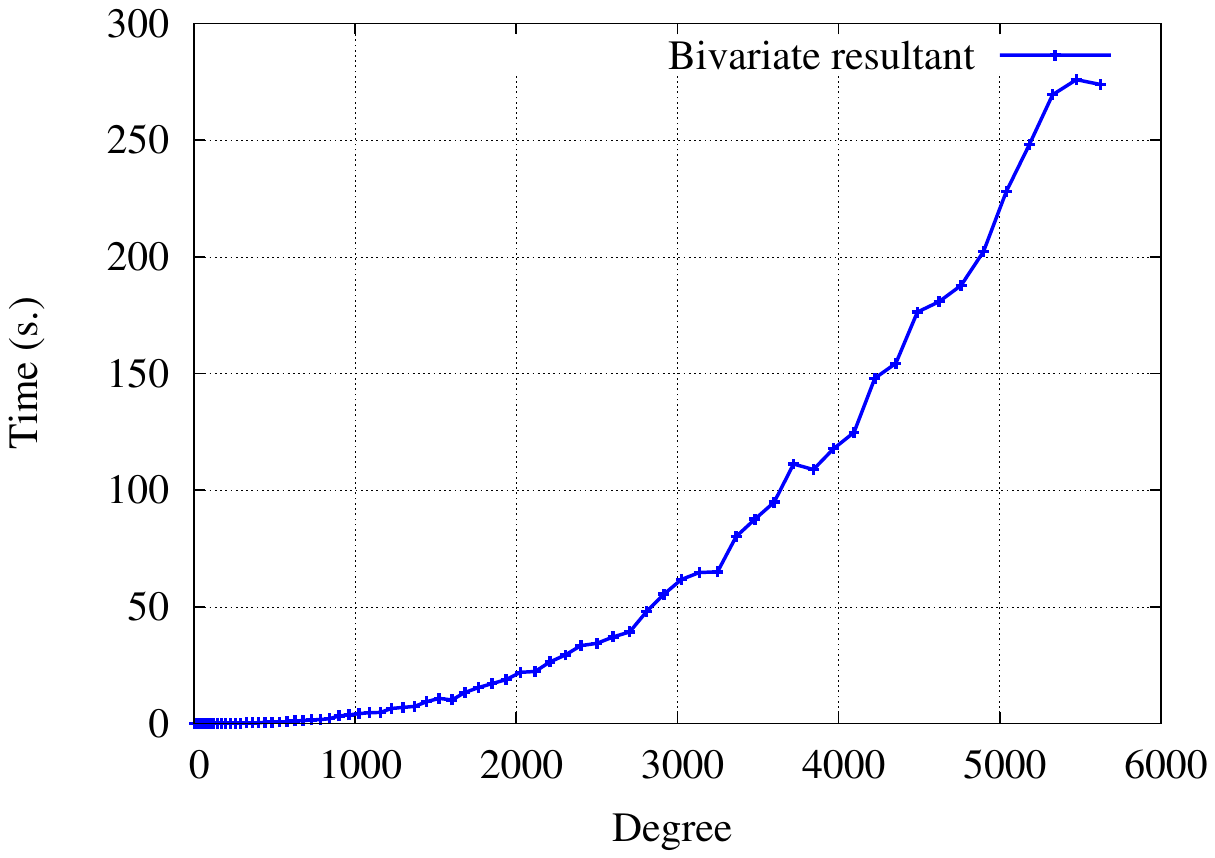}
\else
\includegraphics[width=9cm]{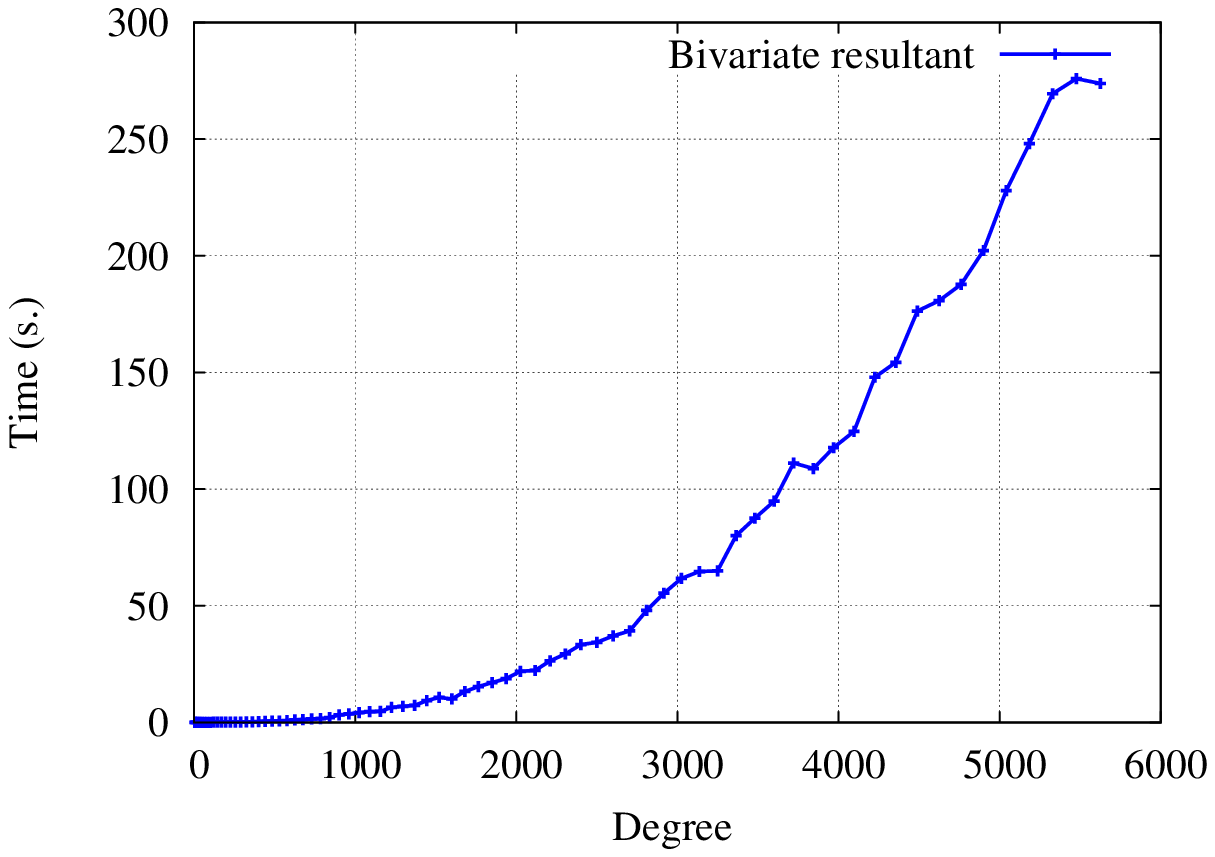}
\fi
\end{center}
\caption{Timings in magma}
\label{fig:magma}
\end{figure}

Figure~\ref{fig:magma} gives timings obtained in Magma, using the
built-in resultant function, on the same set of problems as above. As 
predicted by the complexity analysis, the results are significantly
slower (about two orders of magnitude for the larger problems).

%%%%%%%%%%%%%%%%%%%%%%%%%%%%%%%%%%%%%%%%%%%%%%%%%%%%%%%%%%%%
%%%%%%%%%%%%%%%%%%%%%%%%%%%%%%%%%%%%%%%%%%%%%%%%%%%%%%%%%%%%
%%%%%%%%%%%%%%%%%%%%%%%%%%%%%%%%%%%%%%%%%%%%%%%%%%%%%%%%%%%%

\section{Conclusion}

Several questions remain open after this work. Of course, the most
challenging one remains how to unconditionally get rid of all
exponential factors in multiplication algorithms for triangular sets.
More immediate questions may be the following: at the fine tuning
level, adapting the idea of the Truncated Fourier
Transform~\cite{VdH04} should enable us to reduce the step effect in
the timings of the previous section. Besides, it will be worthwhile to
investigate what other applications can be dealt with using the
``homotopy multiplication'' model, such as the product of matrices with 
entries defined modulo a triangular set, or further tasks such as 
modular  inversion or modular composition.

\bibliographystyle{plain}
\bibliography{mul} 

\begin{thebibliography}{10}

\bibitem{AbOrReVa04}
I.~Abdeljaouad, S.~Orange, G.~Renault, and A.~Valibouze.
\newblock Computation of the decomposition group of a triangular ideal.
\newblock {\em Applicable Algebra in Engineering Communication and Computing},
  15(3-4):279--294, 2004.

\bibitem{AhStUl75}
A.~V. Aho, K.~Steiglitz, and J.~D. Ullman.
\newblock Evaluating polynomials at fixed sets of points.
\newblock {\em SIAM J. Comp.}, 4(4):533--539, 1975.

\bibitem{AuVa00}
P.~Aubry and A.~Valibouze.
\newblock Using {G}alois ideals for computing relative resolvents.
\newblock {\em J. Symb. Comp.}, 30(6):635--651, 2000.

\bibitem{magma}
W.~Bosma, J.~Cannon, and C.~Playoust.
\newblock The {M}agma algebra system. {I}. {T}he user language.
\newblock {\em J. Symb. Comp.}, 24(3-4):235--265, 1997.

\bibitem{BoFlSaSc06}
A.~Bostan, P.~Flajolet, B.~Salvy, and {\'E}.~Schost.
\newblock Fast computation of special resultants.
\newblock {\em J. Symb. Comp.}, 41(1):1--29, 2006.

\bibitem{BoGoPeSc05}
A.~Bostan, L.~Gonz{\'a}lez-Vega, H.~Perdry, and {\'E}.~Schost.
\newblock From {N}ewton sums to coefficients: complexity issues in
  characteristic $p$.
\newblock In {\em MEGA'05}, 2005.

\bibitem{BS:exp}
A.~Bostan and {\'E}.~Schost.
\newblock A simple and fast algorithm for computing exponentials of power
  series.
\newblock Available at \verb|http://algo.inria.fr/bostan/|, 2008.

\bibitem{Brent75}
R.~P. Brent.
\newblock Multiple-precision zero-finding methods and the complexity of
  elementary function evaluation.
\newblock In {\em Analytic computational complexity}, pages 151--176. Academic
  Press, 1976.

\bibitem{CaKa91}
D.~G. Cantor and E.~Kaltofen.
\newblock On fast multiplication of polynomials over arbitrary algebras.
\newblock {\em Acta Informatica}, 28(7):693--701, 1991.

\bibitem{DaMaScWuXi05}
X.~Dahan, M.~Moreno Maza, {\'E}.~Schost, W.~Wu, and Y.~Xie.
\newblock Lifting techniques for triangular decompositions.
\newblock In {\em ISSAC'05}, pages 108--115. ACM, 2005.

\bibitem{DaMoScXi06}
X.~Dahan, M.~Moreno Maza, {\'E}.~Schost, and Y.~Xie.
\newblock On the complexity of the {D}5 principle.
\newblock In {\em Transgressive Computing}, pages 149--168, 2006.

\bibitem{DvTr87}
R.~Dvornicich and C.~Traverso.
\newblock Newton symmetric functions and the arithmetic of algebraically closed
  fields.
\newblock In {\em AAECC-5}, volume 356 of {\em LNCS}, pages 216--224. Springer,
  1989.

\bibitem{FoMo02}
M.~Foursov and M.~{Moreno Maza}.
\newblock On computer-assisted classification of coupled integrable equations.
\newblock {\em J. Symb. Comp.}, 33:647--660, 2002.

\bibitem{Furer07}
M.~F\"{u}rer.
\newblock Faster integer multiplication.
\newblock In {\em 39th Annual ACM Symp. Theory Comp.}, pages 57--66. ACM, 2007.

\bibitem{GaGe99}
{J. von zur} Gathen and J.~Gerhard.
\newblock {\em Modern Computer Algebra}.
\newblock Cambridge University Press, 1999.

\bibitem{GaSc04}
P.~Gaudry and {\'E}.~Schost.
\newblock Construction of secure random curves of genus 2 over prime fields.
\newblock In {\em Eurocrypt'04}, pages 239--256. Springer, 2004.

\bibitem{GaScTh06}
P.~Gaudry, {\'E}.~Schost, and N.~Thi{\'e}ry.
\newblock Evaluation properties of symmetric polynomials.
\newblock {\em International Journal of Algebra and Computation},
  16(3):505--523, 2006.

\bibitem{GoPe04}
L.~Gonz{\'a}lez-Vega and H.~Perdry.
\newblock Computing with {N}ewton sums in small characteristic.
\newblock In {\em EACA'04}, 2004.

\bibitem{KoMo02}
I.~A. Kogan and M.~{{Moreno Maza}}.
\newblock Computation of canonical forms for ternary cubics.
\newblock In {\em {ISSAC'02}}, pages 151--160. ACM, 2002.

\bibitem{Langemyr91}
L.~Langemyr.
\newblock Algorithms for a multiple algebraic extension.
\newblock In {\em Effective methods in algebraic geometry)}, volume~94 of {\em
  Progr. Math.}, pages 235--248. Birkh\"auser, 1991.

\bibitem{LiMoRaSc08b}
X.~Li, M.~{Moreno Maza}, R.~Rasheed, and {\'E}~Schost.
\newblock High-performance symbolic computation in a hybrid
  compiled-interpreted programming environment.
\newblock In {\em ICCSA'08}, pages 331--341. IEEE, 2008.

\bibitem{LiMoSc07}
X.~Li, M.~{Moreno Maza}, and {\'E}.~Schost.
\newblock Fast arithmetic for triangular sets: from theory to practice.
\newblock In {\em ISSAC'07}, pages 269--276. ACM, 2007.

\bibitem{HoMo02}
{M. van Hoeij} and M.~Monagan.
\newblock A modular {GCD} algorithm over number fields presented with multiple
  extensions.
\newblock In {\em ISSAC'02}, pages 109--116. ACM, 2002.

\bibitem{Pan94}
V.~Y. Pan.
\newblock Simple multivariate polynomial multiplication.
\newblock {\em J. Symb. Comp.}, 18(3):183--186, 1994.

\bibitem{Reischert97}
D.~Reischert.
\newblock Asymptotically fast computation of subresultants.
\newblock In {\em ISSAC'97}, pages 233--240. ACM, 1997.

\bibitem{ReYo06}
G.~Renault and K.~Yokoyama.
\newblock A modular algorithm for computing the splitting field of a
  polynomial.
\newblock In {\em Algorithmic Number Theory, ANTS VII}, number 4076 in LNCS,
  pages 124--140. Springer, 2006.

\bibitem{ReVa99}
N.~Rennert and A.~Valibouze.
\newblock Calcul de r\'esolvantes avec les modules de {C}auchy.
\newblock {\em Experimental Mathematics}, 8(4):351--366, 1999.

\bibitem{Schonhage82}
A.~Sch{\"o}nhage.
\newblock The fundamental theorem of algebra in terms of computational
  complexity.
\newblock Technical report, Univ. T{\"u}bingen, 1982.

\bibitem{Schost05}
{\'E}.~Schost.
\newblock Multivariate power series multiplication.
\newblock In {\em ISSAC'05}, pages 293--300. ACM, 2005.

\bibitem{NTL}
V.~Shoup.
\newblock {NTL}: A library for doing number theory.
\newblock \texttt{http://www.shoup.net}.

\bibitem{StMoAm08}
A.~Straub, T.~Amdeberhan, and V.~H. Moll.
\newblock The $p$-adic valuation of $k$-central binomial coefficient, 2008.

\bibitem{Sturmfels93}
B.~Sturmfels.
\newblock {\em Algorithms in invariant theory}.
\newblock Texts and Monographs in Symbolic Computation. Springer-Verlag, 1993.

\bibitem{VdH04}
J.~van~der Hoeven.
\newblock The {T}runcated {F}ourier {T}ransform and applications.
\newblock In {\em ISSAC'04}, pages 290--296. ACM, 2004.

\bibitem{vdH:fnewton}
J.~van~der Hoeven.
\newblock Newton's method and {FFT} trading.
\newblock Technical Report 2006-17, Univ. Paris-Sud, 2006.
\newblock Submitted to J. Symb. Comp.

\end{thebibliography}

\end{document}